\documentclass{article}
\pdfoutput=1
\PassOptionsToPackage{numbers, compress}{natbib}

\usepackage{style}
\usepackage[utf8]{inputenc} 
\usepackage[T1]{fontenc}    
\usepackage{url}            
\usepackage{booktabs}       
\usepackage{amsfonts}       
\usepackage{nicefrac}       
\usepackage{microtype}      
\usepackage{subcaption}     
\usepackage{graphicx}

\usepackage{multirow}
\usepackage{footnote}
\usepackage{amsmath}
\usepackage{dsfont}
\usepackage{mathtools}
\usepackage[colorlinks]{hyperref}
\usepackage[svgnames]{xcolor}
\usepackage{adjustbox}
\AtBeginDocument{
\hypersetup{
citecolor=MidnightBlue,
linkcolor=Brown,   
urlcolor=MidnightBlue}
}      

\makeatletter
\newcommand{\printfnsymbol}[1]{%
  \textsuperscript{\@fnsymbol{#1}}%
}
\makeatother

\title{Wisdom of the Crowds or Ignorance of the Masses? A data-driven guide to WSB}

\author{%
  Valentina Semenova\thanks{Equal contribution}\\
  University of Oxford, INET, \\
  Oxford-Man Institute for Quantitative Finance\\
  \texttt{semenova@maths.ox.ac.uk}
  \And
  Dragos Gorduza\printfnsymbol{1}\\
  University of Oxford, \\ Oxford-Man Institute for Quantitative Finance\\
  \texttt{dragos.gorduza@st-annes.ox.ac.uk}
  \And
    William Wildi\\
  University of Oxford, \\ Oxford-Man Institute for Quantitative Finance\\
  \texttt{william.wildi@wolfson.ox.ac.uk}\\ 
  \And
  Xiaowen Dong\\
  University of Oxford, \\ Oxford-Man Institute for Quantitative Finance\\
  \texttt{xiaowen.dong@eng.ox.ac.uk}
    \And
  Stefan Zohren\\
  University of Oxford, \\ Oxford-Man Institute for Quantitative Finance\\
  \texttt{stefan.zohren@eng.ox.ac.uk}
}

\begin{document}

\maketitle

\begin{abstract}

    A trite yet fundamental question in economics is: What causes large asset price fluctuations? A tenfold rise in the price of GameStop equity, between the 22\textsuperscript{nd} and 28\textsuperscript{th} of January 2021, demonstrated that herding behaviour among retail investors is an important contributing factor. This paper presents a data-driven guide to the forum that started the hype -- WallStreetBets (WSB). Our initial experiments decompose the forum using a large language topic model and network tools. The topic model describes the evolution of the forum over time and shows the persistence of certain topics (such as the market / S\&P500 discussion), and the sporadic interest in others, such as COVID or crude oil. Network analysis allows us to decompose the landscape of retail investors into clusters based on their posting and discussion habits; several large, correlated asset discussion clusters emerge, surrounded by smaller, niche ones. A second set of experiments assesses the impact that WSB discussions have had on the market. We show that forum activity has a Granger-causal relationship with the returns of several assets, some of which are now commonly classified as `meme stocks', while others have gone under the radar. The paper extracts a set of short-term trade signals from posts and long-term (monthly and weekly) trade signals from forum dynamics, and considers their predictive power at different time horizons. In addition to the analysis, the paper presents the dataset, as well as an interactive dashboard, in order to promote further research. 

\end{abstract}
\begin{itemize}
    \item The paper uncovers text-based and user pattern-based asset clusters within the WSB forum by applying large language models and network techniques to WSB submission and comment data.
    \item We explore the reaction of WSB users to market moves through an analysis of cumulative abnormal returns preceding and following submissions, and observe that users are generally reactive to price and that this pattern is particularly distinctive in `meme' stocks.
    \item We quantify the impact that WSB activity has had on the market through documenting whether posts on the forum Granger-cause asset returns, and through studying the predictive signals within `due diligence' posts. 
    \item In addition to the analysis, the paper presents the dataset of hand-annotated due diligence posts, posts labeled with a sentiment classifier, as well as an interactive dashboard, to promote further exploration and research. 
\end{itemize}

\newpage
\section{Introduction}
\label{sec:intro}

    Social media has changed the structure of our society. As many as 4.9 billion people, or 61\% of the world population, are active social media users, each just a few clicks away from the next viral phenomenon. People turn to their accounts for everything from news to product suggestions. Now, a growing audience turns to social media for promising stock market gambles. Even though investor discussion forums have existed for decades, the \url{r/WallStreetBets} (WSB) subreddit was arguably the first to reach an unprecedented retail following -- since it's creation in 2012, the forum grew exponentially in membership, attracting followers not only through lucrative trade ideas, but also through the promise that coordination among smaller retail traders could unseat investment giants. Their aspirations became a reality in January 2021 when the forum's cherished stock, GameStop (GME), experienced a 22-fold rise in asset price, while \textit{Melvin Capital}, an investment fund with a large short position in GME, experienced a 30\% decline in its value.

    A single trader with \$1,000 in her bank account couldn't move the markets. 750 thousand individuals on the other hand, each with \$1,000 in their bank accounts, could have bought all of the floating shares of GameStop at its early January price of \$17.25 per share. However, how did they overcome the colossal coordination challenge? What was the structure of the discussion on \url{r/WallStreetBets} and how did the conversation evolve? In this paper, we dive under the hood with the goal of shedding light on the dynamics of the forum which started the hype.
    
    Even though the literature has explored specific phenomena within the WSB forum \citep{boylston2021wallstreetbets,semenova2021reddit,Buz2021wsb,witts2021irrational,chacon2022will}, we believe that our work is the first to take a broad approach. Our goal with this paper is to thoroughly characterize the trading signals and behaviours on WallStreetBets. We focus on the following questions:
    \begin{itemize}
        \item What are the primary topics of interest to WSB users, and how do they relate to each other in the forum discussion landscape?
        \item What assets are forum participants discussing, and can their conversations be used as indicators of market moves?
        \item Are WSB users trend-followers or predictors of asset returns?
    \end{itemize}
    We answer these questions by applying several machine learning tools to the dataset of WallStreetBets submissions and comments:
    \begin{enumerate}
        \item We extract text features from the forum data, such as sentiments and topics using large language models \cite{araci2019finbert,angelov2020top2vec}, as well as tickers mentioned within posts.
        \item We estimate the relationship between assets using a network approach applied to both our extracted topics, and user submissions. 
        \item We approximate the interaction of asset returns and user sentiment by estimating the cumulative abnormal returns in the days preceding and following submissions. 
        \item We study whether WSB discussions forecast market returns by estimating a Granger causal relationship between the two.
    \end{enumerate}

    The text features within our data allow us to uncover several important characteristics. Our topic model shows how the interests of the forum have evolved over time. Certain topics, such as the discussion of Tesla and Elon Musk, retain the interest of WSB users. Others, such as oil and or COVID, briefly grab the forum's attention, but are promptly replaced by other discussion themes. Our sentiment model allows us to extract a perspective taken by users (bullish, bearish, or neutral) about the future movement of the asset mentioned in a submission. This sentiment measure, in turn, allows us to understand the predictive power of the forum and user reactions to market moves.

    The network approach allows us to cluster assets based on: i) whether they are frequently mentioned together while discussing the same topic (\textit{Topic Network}), and ii) whether the same group of users are interested in both assets (\textit{Submission Network}). We observe that distinct clusters of similar assets emerge. Posts within the large asset clusters generally tend to incorrectly forecast future asset prices. However, smaller, niche clusters have better market prediction performance, perhaps indicating that close-knit groups of well-informed retail investors may have market insights.

    We analyze the seven-day Cumulative Abnormal Returns (CAR) in the fourteen days preceding and following a post about an asset. The CAR is the asset log-return, which is distinct from simultaneous market returns, and is typically used to model sudden price shocks and event detection in financial markets. We observe that, overall, WSB appears to be reactive to, rather than predictive of, CAR in an asset. The average CAR appears to run up gradually immediately before a post, and subsequently experiences a sharp decline. The pattern is particularly distinctive in `meme' stocks, where high asset returns potentially reinforce the hype. In stocks with a broader following, such as MSFT or AAPL, CAR appears flat around WSB posts. 

    In a final exercise, we study a Granger causal relationship between asset returns and sentiments in posts. A time series A is said to Granger-cause B if it can be shown that the \textit{lagged} values of A provide statistically significant information about \textit{future} values of B. Our analysis highlights that a Granger causal relationship exists between the sentiments on WSB of many `meme' assets and their returns -- the result is not surprising, since conventional wisdom is that the WSB discussions heavily influenced `meme' stocks. The Granger causal analysis allows us to look at the comparative degree of influence between different tickers, and also to identify new assets which may have been influenced by WSB but have gone under the radar - such as SnapChat.

    Beyond the immediate quantitative insights, an additional goal of the paper is to provide tools and data for researchers and practitioners. We hope that this research will contribute to a broader effort within the financial industry to take advantage of and evaluate tradeoffs in non-stationary and quickly-evolving signals. We make the dataset used within the paper, manual annotations, as well as code database, public. Furthermore, in an effort to encourage a more interactive data experience for practitioners, we publish our insights in a dashboard, allowing users to explore various aspects of the dataset and evaluate opportunities for market returns.\footnote{\url{https://sites.google.com/view/wsbtrialsite?usp=sharing}}

    \paragraph{Relevant Literature} We believe that this paper may be of interest to those looking into social media in finance and the WallStreetBets forum specifically, as well as to those studying new, noisy signals within the finance community. 

    Our paper takes inspiration from an emerging literature looking at the relationship between social media activity and stock market prices. It is well understood that new information about a company (such as from a news report) can affect its stock price, as well as the stock price of similar or competing companies \cite{Wan2021}. Social media data provides information beyond news: shedding light on how individuals engage with financial information, which can reflect the sentiment of the market more broadly. Many researchers have exploited this to find associations between publicly expressed sentiment on Twitter /StockTwits and stock market prices \cite{TwitterSentiment2, PagoluTwitterSent, TwitterSentiment1, Duz2021,azar2016wisdom,agrawal2018momentum}. Social media has thereby provided rich new text data for quantitative finance researchers to exploit, and is now a key input (along with news articles and company reports) for text analysis in finance (see \cite{gao2021review} for a review). We contribute to this literature both through our analysis of a specific social media dataset, the WallStreetBets forum, and through providing annotated data for future research. A more targeted set of papers focus on specific characteristics of the forum. We add to this literature by taking a more broad approach to analyzing the forum, and discuss recent works in Appendix \ref{app:lit_review}.

\section{Data}
\label{sec:data}

\paragraph{Data collection} Submission and comment data from WSB is collected through the Pushshift Reddit database \cite{baumgartner2020pushshift}, which provides an archive of all posts and comments on Reddit. We collect data from the inception of the forum in April 2012 up to 24 June 2022. Posts on WSB typically refer to companies by their stock symbol, and we use the work of \cite{Semenova21} to obtain a list of stock symbols (from Yahoo Finance and Compustat) and extract these from the submission text. Stock price data is sourced from from \href{https://www.alphavantage.co/}{AlphaVantage}.

\paragraph{Due diligence posts} Due diligence (DD) posts have attracted attention in the literature as `higher quality' posts from WSB, which contain meaningful stock-level insights. To obtain the subset of due diligence (DD) reports in the forum, we firstly filter to posts that have the DD flair. We supplement the posts containing the DD flair with posts that contain the DD acronym. We manually review every single potential DD post (i.e. flaired posts that are not removed by moderators, or contain the text `DD'), and we remove any post that does not appear to be a valid DD -- i.e. does not contain market insight. The DD posts are sufficiently small in number that we can manually label each post as bullish/positive or bearish/negative manually (neutral posts are removed). We find 77 percent of posts are bullish, and 23 percent are bearish. 

\paragraph{Sentiment classification} For the forum as a whole there are far too many posts for manual classification. Previous papers have either relied on counting keywords, or they have used off-the-shelf rule based sentiment classifiers, resulting in certain mis-classification issues, discussed in Appendix \ref{app:data}. To resolve these issues, we firstly hand-code a random sample of 4,000 WSB posts, categorising them as bullish, bearish, or neutral depending on what the author's attitude is on future price increases. Of the posts in the sample 41 percent are bullish, 37 percent are neutral and 22 percent are bearish.\footnote{Since bearish posts tend to be a minority on the forum, bearish posts were up-sampled to reduce class imbalance.} We then take a pre-trained BERT \cite{devlin2018bert} model called FinBERT \cite{araci2019finbert}, which we fine-tune using our manually labelled sample. Our final model had a test accuracy of 69.2 percent, which we consider sufficient due to the challenging task of predicting future-facing sentiment. 

Further details on data sourcing, cleaning and annotation are presented in Appendix \ref{app:data}.

\section{WallStreetBets Forum Dynamics}
\label{sec:forum_dynamics}

In this section, we characterise the discussion landscape of the WSB forum. We consider how assets are related on WSB in terms of topic and user interest. Additionally, we explore the relationship between WSB and markets through studying the returns preceding and following posts. The section that follows discusses predicting market movements using data from WSB. 

\subsection{The topic landscape of WSB}
\label{subsec:topic_model_data}

    \begin{figure}[ht!]
    \begin{center}
        \begin{subfigure}[t]{\textwidth}
            \includegraphics[width=\textwidth]{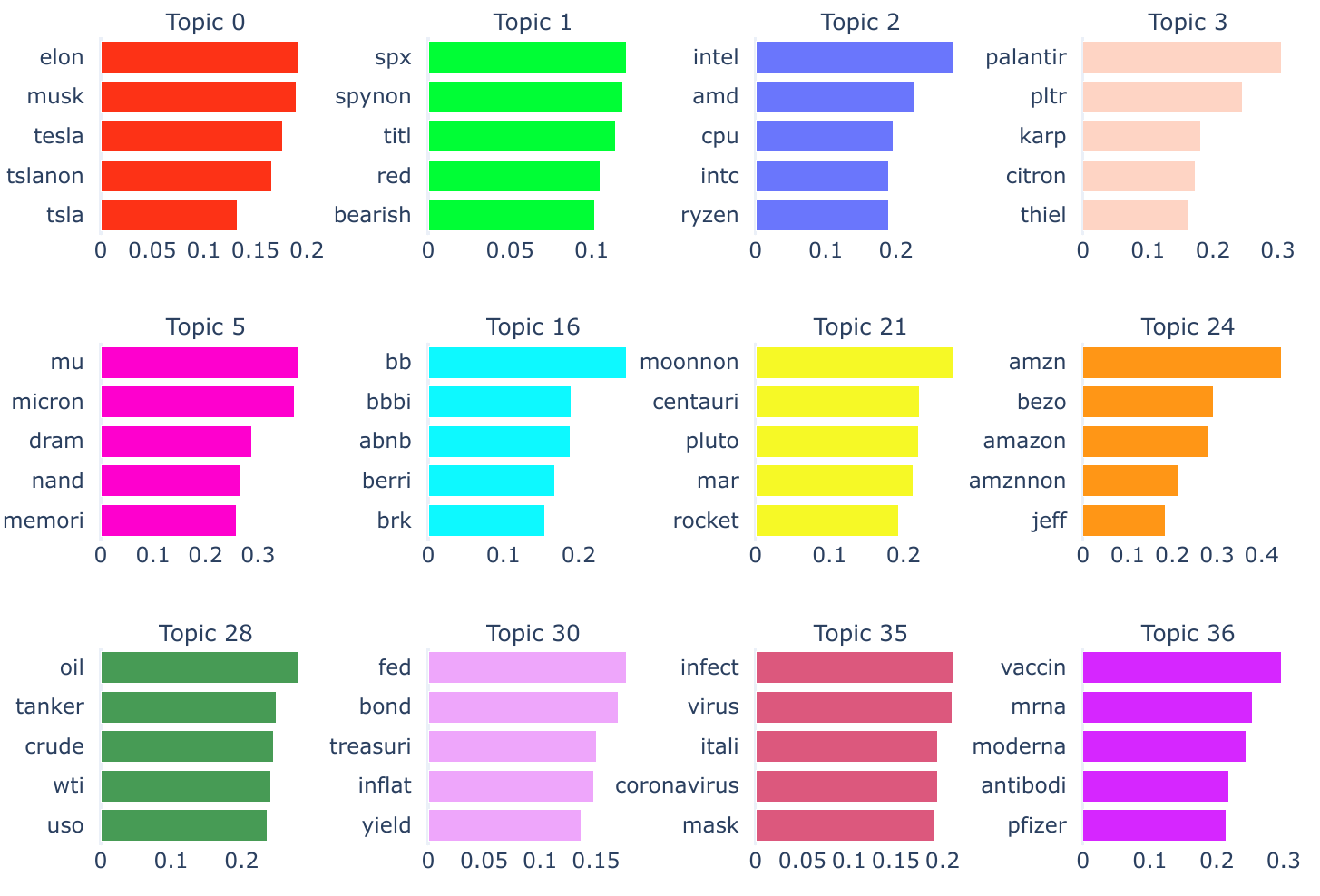}
                    \caption{Sample of Representative Topics Discussed on WSB; the figure displays topics and the frequency with which the most popular words appear within the topics (ranked by the categorical term frequency - inverse document frequency score \cite{angelov2020top2vec}). }
        \label{fig:represented-topics}
        \end{subfigure}
        \begin{subfigure}[t]{\textwidth}
            \includegraphics[width=\textwidth]{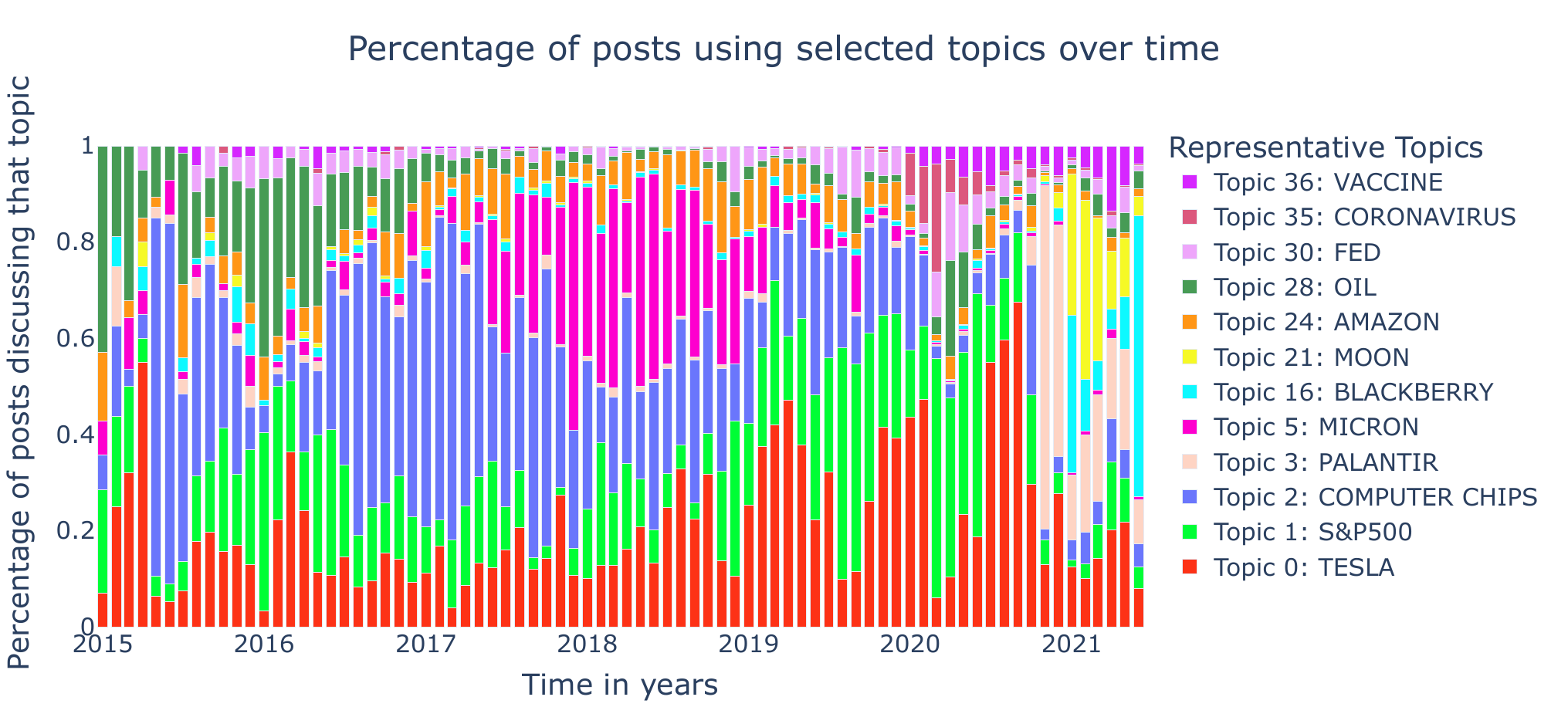}
        \caption{Topic Distribution over Time}
        \label{fig:top_30_topics_overtime}
        \end{subfigure} 
        \vspace{-0.8em}
        \caption{The WallStreetBets Discussion; we present several representative topics from WSB in Figure \ref{fig:represented-topics}, as well as their relative importance within the forum in Figure \ref{fig:top_30_topics_overtime}. }
        \label{fig:topic_model}
        \end{center}
    \end{figure}
    
Topic models are often employed to map out the distribution of information in a textual corpus \citep{blei2003latent,angelov2020top2vec}. In economics and finance, they are used to uncover the latent discursive directions that individuals can follow when expressing an opinion, which allows researchers to quickly detect salient points in the discourse and use them as signals for downstream analysis \cite{gentzkow2019text}. However, prior work leverages older topic models which do not make full use of the new modelling capacities offered by large language models (LLMs) \citep{schou2022we}. We use the Bert LLM Topic package to perform topic modelling on WSB, which accounts for non-linear semantic similarities between texts which go beyond word-level co-occurrences (a limitation of traditional topic models) \citep{angelov2020top2vec}. This model gives us a mapping from each post in the sample to a learnt set of topics. Each topic has a unique set of representative terms associated with it, and posts are mapped to topics based on how well the text within the post matches that of a particular topic.

Figure \ref{fig:topic_model} highlights our key findings from employing the BERT topic model: Figure \ref{fig:represented-topics} present some of the key topics of discussion on WSB, while Figure \ref{fig:top_30_topics_overtime} displays their prevalence on the forum over time. Each post within WSB is given a probability vector whose length is equal to the number of extracted topics, identifying the extent to which the post is associated with different topics. For example, as post which discusses pharmaceutical companies and how they will profit from the COVID vaccine will be given a higher probability of being associated with topics 35 and 36. 

Figure \ref{fig:top_30_topics_overtime} is constructed by first splitting our dataset into months and then calculating the fraction of the discourse dedicated to that topic within that month for the ten most popular topics. The span of topics covered here is a subset of the overall topics identified by the model, which serve to illustrate the key areas of attention for the forum across time. We note that several topics and their prevalence follow news, such as CORONAVIRUS (topic 35) which appears in March 2021 and VACCINE (topic 36), which are more prevalent during the introductions of the first COVID-19 vaccines in end of the year 2021. The topic model also tracks the emerging focus of the forum on so called `meme stocks' such as TESLA (topic 0), PALANTIR (topic 3), and BLACKBERRY (topic 16). Contrary to other transient topics, we detect a persistent interest of the forum in SPY ( S\&P500 ) as noted by the prevalence of topic 1, which represents attention toward the overall state of the market.  

\subsection{How are assets related to each other on WSB?}

    \begin{figure}[ht!]
    \begin{center}
        \begin{subfigure}[t]{\textwidth}
        \caption{\textit{Topic Network} construction illustration}
            \includegraphics[width=\textwidth]{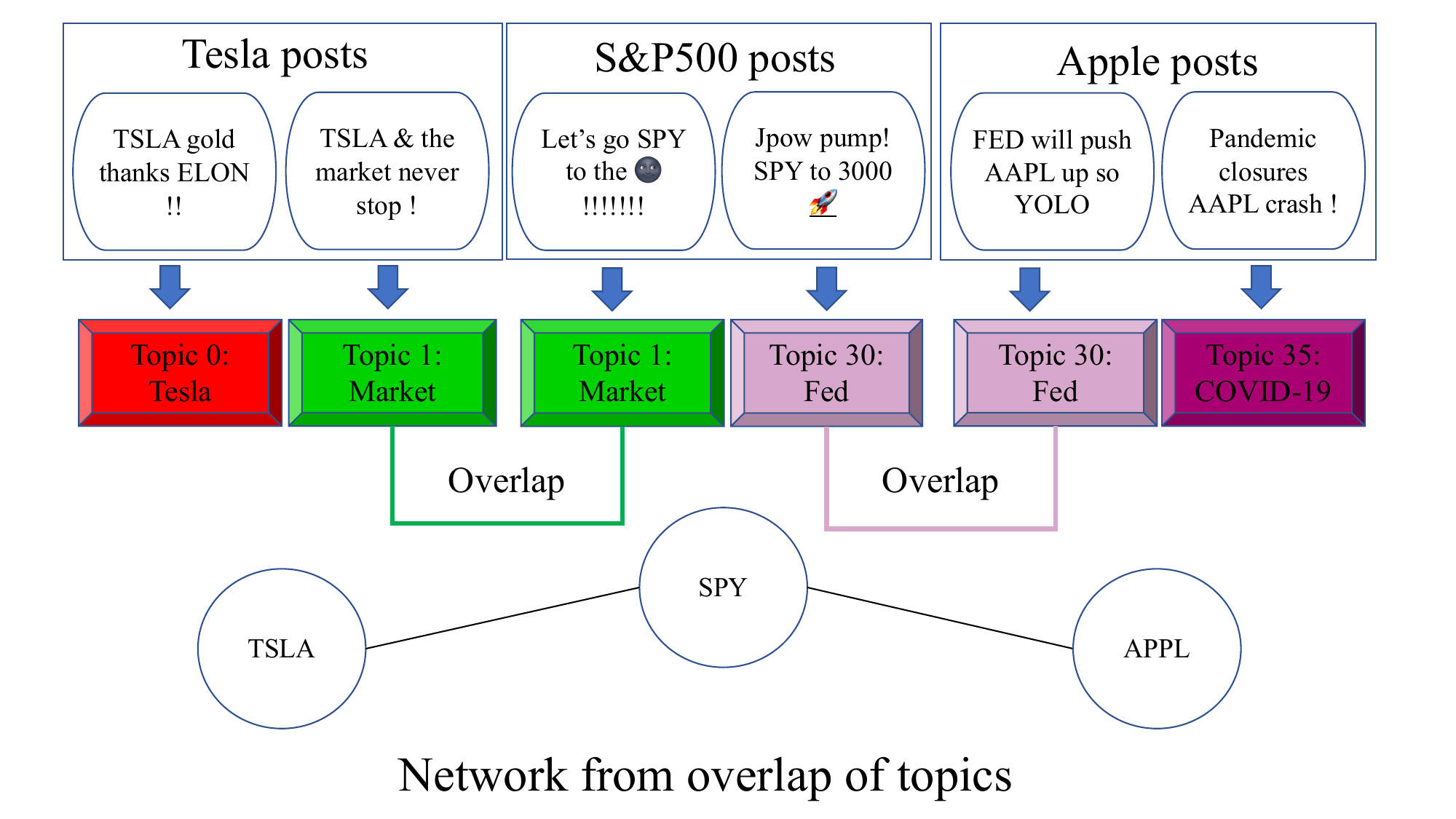}
        \label{fig:topic_network_construction}
        \end{subfigure}
        \begin{subfigure}[t]{\textwidth}
            \caption{\textit{Submission Network} construction illustration}
            \includegraphics[width=\textwidth]{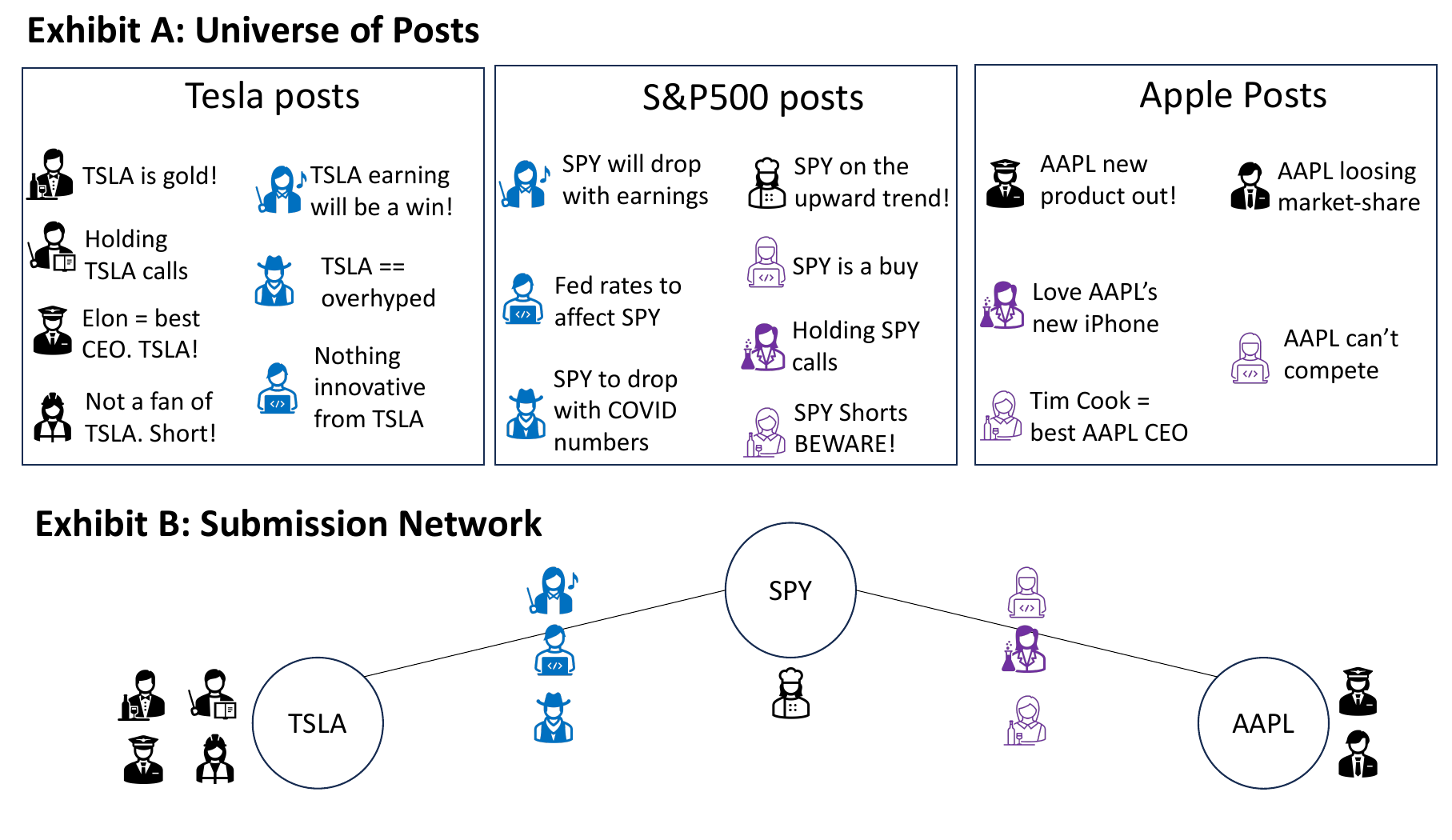}
        \label{fig:sub_network_construction}
        \end{subfigure} 
        \vspace{-0.8em}
        \caption{Network Construction; we demonstrate how links are identified between assets in the \textit{Topic Network} and \textit{Submission Network}.}
        \label{fig:network_construction}
        \end{center}
    \end{figure}

The WSB forum is uniquely suited to study the relationships between assets, as they are perceived and discussed together by retail investors. We choose two methods to map the relationships between assets and to create a ticker-to-ticker network structure: the \textit{Topic Network} approach and the \textit{Submission Network} approach. The \textit{Topic Network} uses the frequency with which assets are discussed within the same topics to create inter-asset connections, indicating that investors perceive the assets to be influenced by the same general financial trends. The \textit{Submission Network} tracks which groups of investors are interested in, and create submissions about, the same assets. We describe both approaches below.

\paragraph{Topic Network} In the \textit{Topic Network} approach, we link assets based on the frequency with which they are mentioned in the same set of topics. The intuition behind the approach rests on the idea that assets that are discussed within the same financial topics are likely related to each other through some underlying fundamental relationship. 

We explain our intuition through a simplified example. Let us consider two assets: i) interest rate swaps and ii) bonds. We extract a topic from our topic model about the FED, and we observe that both bonds and interest rate swaps are frequently brought up in posts that are labeled with the `FED' topic. This is unsurprising, given the fact that FED decisions would strongly affect the valuations of both bonds and interest rates swaps. In our \textit{Topic Network} exercise we would place a connection between bonds and interest rate swaps, since they are linked by the `FED' topic. This intuition can be extended to better understand the meaning behind the tickers linked through out topic model: they are tickers that are frequently brought up under the same topics, indicating that they are linked through some underlying economic discussion theme. 

Figure \ref{fig:topic_network_construction} provides a toy example. TSLA and SPY are connected because they are both mentioned in the topic discussing the overall stability of the market (\textit{market} topic); SPY and APPL are connected since they are both discussed in conjunction with the \textit{FED} topic.

\paragraph{Submission Network} The community structure within WSB offers a different perspective on how assets are related within the forum. To construct our submission network, we look at the overlap of users who create posts about two assets. Two assets are linked if there is a sufficiently large fraction of users discussing both assets simultaneously. 

Figure \ref{fig:sub_network_construction} provides a simplified example. We observe a subset of users creating posts about TSLA and SPY simultaneously, and a different group of users writing about AAPL and SPY. However, we do not observe the same users creating posts about TSLA and AAPL. Therefore, we create two links: TSLA / SPY  and AAPL / SPY. 

In practice, our network construction exercise is slightly more complicated. We create weighted links between assets, normalized by the total number of users mentioning each asset. This means that if asset A is discussed by ten users, and all of them also post about asset B - we would create a link with weight one from asset A to B. However, if 10,000 users posts about asset B, we would not create a link back from B to A, since only a negligible fraction of posters about B care about A. Therefore, the \textit{total weight} of the link between A and B would be 1 from the first created link. In practice, our threshold for including a link in our weight calculation is 20\% -- at least 20\% of users posting about one ticker must be posting about the other ticker. Relating this back to Figure \ref{fig:sub_network_construction}, we observe that three out of seven SPY posters also post about AAPL (weight 3/7) and that three out of five AAPL posters post about SPY (weight 3/5) -- the total weight for the link between AAPL and SPY would, therefore, be 3/5+3/7. We filter our submission network to contain tickers that are mentioned at least 150 times on the forum.

\subsection{Asset network structures}

    \begin{figure}[ht!]
    \begin{center}
        \begin{subfigure}[t]{\textwidth}
        \caption{Topic network and extracted clusters}
            \includegraphics[width=\textwidth]{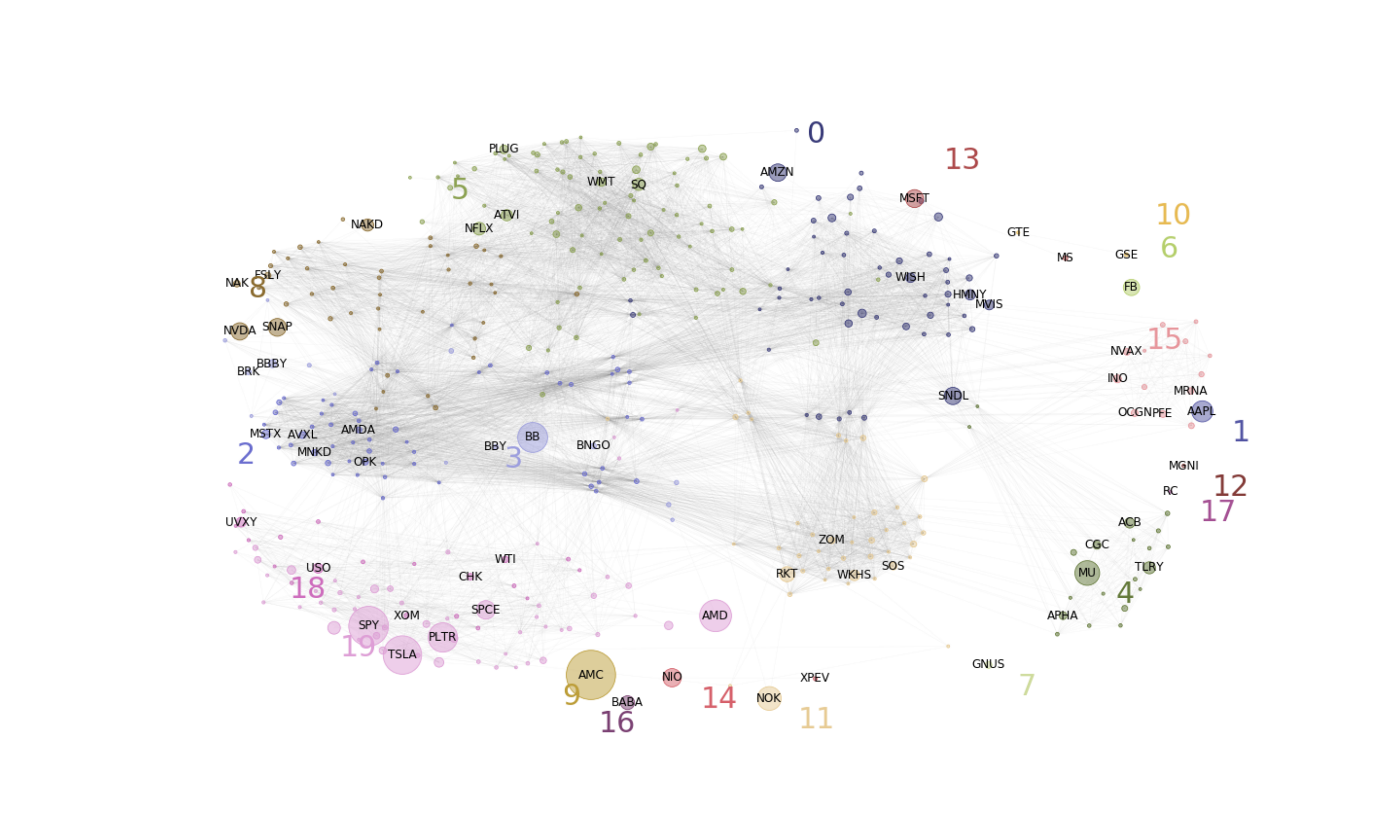}
        \label{fig:topic_network_observed}
        \end{subfigure}
        \begin{subfigure}[t]{\textwidth}
            \caption{Submission network and extracted clusters}
            \includegraphics[width=\textwidth]{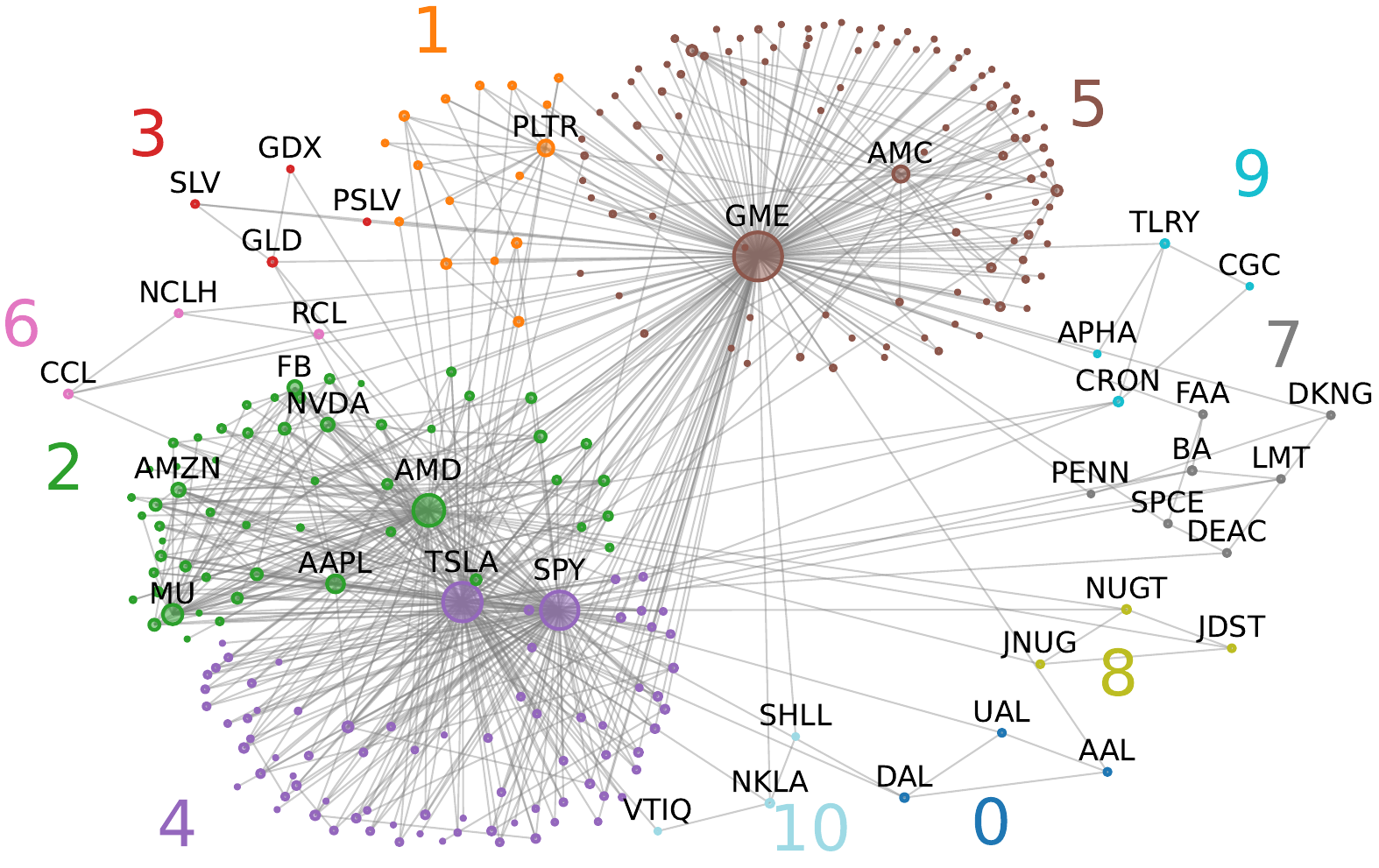}
        \label{fig:sub_network_observed}
        \end{subfigure} 
        \vspace{-0.8em}
        \caption{Extracted Network from WSB; we illustrate the ticker networks extracted using both the \textit{Topic Network} and \textit{Submission Network} approaches. Clusters of closely connected tickers are extracted using the Leiden algorithm \citep{traag2019louvain} and highlighted with the same color in both networks. In Figure \ref{fig:topic_network_observed}, ticker nodes are scaled by the relative size of their number of posts; only the top 50 most mentioned topics are displayed. In Figure \ref{fig:sub_network_observed}, nodes are scaled by the number of connections they have. }
        \label{fig:networks_observed}
        \end{center}
    \end{figure}


\paragraph{Topic Network -- Results} Figure \ref{fig:topic_network_observed} presents our \textit{Topic Network} -- links indicate which two tickers are likely to be mentioned together within the same economic discussion topics (a complete list of tickers and their associated clusters is presented in Appendix \ref{app:topic_model}). We observe that larger and frequently discussed companies, such as AMC, have more cohesive topics which are not referenced in many other companies' posts. Therefore, the posts of several tickers are isolated from the main connected component of the topic network. These includes: AAPL (cluster 1), FB (cluster 6) AMC (cluster 9), and BABA (cluster 16). We interpret the isolation of these network components as the result of a systematic use of a \textit{limited} and \textit{distinct} set of topics when discussing these stocks. A post about AAPL will mobilise only topics similar to other AAPL discussions: focusing the attention of readers more narrowly on discussion points unique to Apple Inc as opposed to the wider market. Contrary to that, smaller companies tend to have a more central position in this network, and are often connected to stocks with higher market caps. This is visible by the more spread out cluster 19, which encompasses a variety of both smaller and larger companies as well as ETFs, likely connected through broader discussions about the economic climate. 

Additionally, we find smaller clusters of assets related to a narrow set of economic themes within the market. A prime example of this is cluster 15, including NVAX, PFE, MRNA. These firms are likely linked by the same topics of drug discovery, FDA approvals and the COVID vaccine, in a way that is distinct to this particular group of assets.

We use our \textit{Topic Network} as a distance mapping between firms that takes into account their `discursive environments'. We interpret tickers belonging to the same cluster as an indication of greater similarity within their discussion topics. Consequently, belonging to the same cluster could be an indicator of a more correlated information set driving returns for all assets within the cluster.

\paragraph{Submission Network -- Results} Figure \ref{fig:sub_network_observed} presents the results of a clustering exercise on the \textit{Submission Network} \citep{traag2019louvain}. We observe that similar assets appear to be mentioned within the detected clusters -- implying that individuals self-select into discussions about certain asset categories. Said differently, a distinct group of investors is interested in marijuana stocks to those discussing the cruise line industry. This is perhaps most pronounced in the small cluster of gold ETFs containing JNUG, NUGT, JDST -- cluster 8. Several other smaller clusters, dominated by a niche sector, are also visible in clusters 0, 3, 8, 9. We observe a large cluster around the SPY ETF and TSLA (cluster 4), as well as a large `meme stock' clusters around GME (cluster 5). We observe a distinct cluster with some large, popular tech stocks including AAPL, FB, MU, AMZN, AMD (cluster 2). In Appendix \ref{app:clusters_returns}, we explore the correlation among assets in different clusters, as well as returns to WSB advice within the different clusters. 

\paragraph{Returns and Correlations across Clusters}

In our \textit{Submission Network}, we observe that distinct groups of investors are interested in different types of tickers: some users come to WSB in order to discuss pharmaceutical companies, while others are interested in trading natural resource ETFs, or airline stocks. For this reason, tickers clustered through the \textit{Submission Network} tend to have returns that are highly correlated -- clustered assets appear to have an average  return correlation of greater than 0.2, while assets across our entire dataset appear to exhibit an average correlation of 0.16. The most highly correlated asset returns appear to be within clusters 0, 6, and 9: the returns of the assets within these clusters exhibit correlations of 0.80, 0.84 and 0.60, respectively. The average next-day returns of investing according to the extracted sentiments within submissions is statistically significant and negative across most clusters. The most negative average returns are exhibited within clusters 0 and 5; average returns of submissions in cluster 7, on the other hand, are statistically significant and positive. This demonstrates that most investors on WSB lack insights into future market moves. Notably, investors into the `hype' cluster tend to lose the most. However, niche groups of investors may have worthwhile market insights -- as demonstrated by returns for cluster 7. Appendix \ref{app:clusters_returns} presents returns and correlations across clusters. 

A similar pattern can be observed from the \textit{Topic Network} clusters. Investing according to the sentiments of submissions within most topic clusters results in a statistically significant, negative average next day return. However, similarly to our observations from the \textit{Submission Network}, certain topic clusters exhibit positive daily returns, on average: cluster 1 (AAPL), 13 containing MSFT and MS, 14 containing NIO and XPEV (two electric car makers), and 16 containing BABA. We notice that smaller clusters with fewer assets perform better than much larger clusters, again demonstrating that the forum as a whole may lack market insights, however specific topics and groups of investors may have promising insights.


\subsection{Returns preceding and following posts}
\label{subsec:car_analysis}
What are the characteristics of assets returns before and after they are mentioned on WSB? Is there any evidence that WSB users can predict returns, or are they trend followers just like other retail investors? 

\paragraph{Framework} We consider how asset prices are changing shortly before and after posts on the forum. To do this, we follow the methodology of \cite{Wan2021}, which analyzes market movements around news events by looking at changes in abnormal return (AR). AR is derived from the Capital Asset Pricing Model (CAPM) \cite{capm}, which describes returns for company $i$ at time $t$ as follows:
\begin{equation}
    r_{i,t} = \alpha_t + \beta_t r_{m,t} + \epsilon_{i,t}.
\end{equation}
Here $r_{i,t}$ is the log return in the price of stock $i$ on day $t$ compared with the previous day, where $r_{i,t} = \log(\frac{p_{i,t}}{p_{i, t-1}})$ and $p_{i,t}$ is the adjusted close stock price. $r_{m,t}$ is the return of the market (in our case, we use the S\&P 500), hence $\beta$ captures stock price moves that are driven by movements in the wider market. $\alpha_{i,t}$ captures stock over/under-performance relative to the market. $\epsilon_{i,t}$ is a stochastic error term, often referred to as abnormal return (AR).

AR tends to have high magnitude in the presence of sudden price shocks (for example, a news story about a particular company), and hence it is often used for event detection in financial markets. In our case, it is useful for assessing if WSB sentiment is able to provide indication about a future price shock. We fit the CAPM model to each stock and day in our data, using a moving 180 day window. Following \cite{Wan2021}, we calculate the \textit{seven-day cumulative abnormal return} (CAR) in stock $i$ preceding day $t$:
\begin{equation}
	CAR_{i,t} = \sum_{t-6}^{t} \epsilon_{i,t}.
\end{equation}
Once we round the timestamp of a WSB submission to the nearest future market close time, we can match the CAR for each company/time to the WSB data, along with a time series of how the CAR changed over the 14 trading days preceding and following a WSB post. 

\begin{figure}[ht!]
\begin{center}
        \begin{subfigure}[t]{.49\textwidth}
             \includegraphics[width=\textwidth]{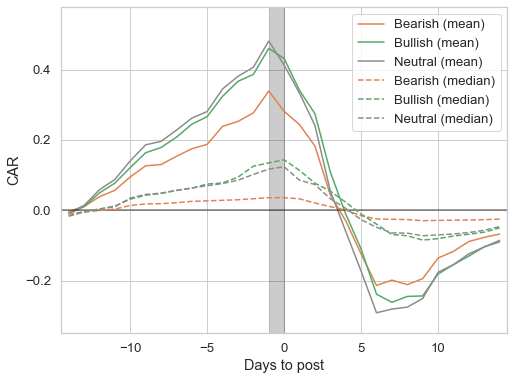}
            \subcaption{CAR for all posts}
            \label{fig:CAR_all}
        \end{subfigure}
        \begin{subfigure}[t]{.49\textwidth}
            \includegraphics[width = \textwidth]{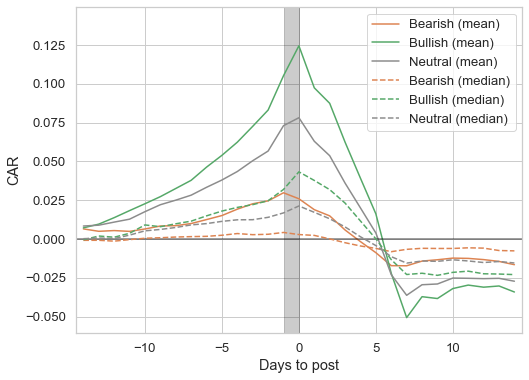}
            \subcaption{CAR for all posts, excluding GME and AMC}
            \label{fig:CAR_excluding}
        \end{subfigure}  
\caption{Average seven-day cumulative abnormal return 14 trading days before and after post submission, grouped by post sentiment. The black bar shows the window of time when posts were written. Figure \ref{fig:CAR_all} uses all posts in the data, while Figure \ref{fig:CAR_excluding} excludes GME and AMC.}
\label{fig:CAR_main}
\end{center}
\end{figure}

\paragraph{CAR Results}
Figure \ref{fig:CAR_main} shows how the average seven-day cumulative abnormal return (CAR) changes 14 trading days before and after a submission. Figure \ref{fig:CAR_all} shows the CAR plots for all WSB posts, while Figure \ref{fig:CAR_excluding} excludes GME and AMC, as the short squeeze events on these stocks created exceptionally high abnormal returns. We observe the following: firstly, CAR tends to be rising up to the submission date, and rapidly declines afterwards. This suggests that WSB activity tends to follow significant price changes in the market, rather than providing a leading indicator of price changes. Secondly, the shape of the curves are quite similar, regardless of sentiment breakdown -- it is primarily the magnitude of the CAR that differs. This suggests that high CAR is associated with more posts of any sentiment, although the effect is more pronounced for bullish posts.

One might attempt to infer from the sharp fall in CAR after posts that shorting stocks that are popular on WSB might be a profitable trading strategy. This inference is, however, mistaken for two reasons. Firstly, negative CAR does not necessarily imply negative returns. Secondly, and more importantly, the CAR plots suggest that WSB sentiment peaks when CAR peaks, however, predicting when sentiment will peak is challenging. If sentiment is already high, and AR increases, then sentiment is likely to continue to increase. In Appendix \ref{app:CAR}, we consider a breakdown of the CAR plots for the twenty most popular stocks on WSB and show that the pattern presented in Figure \ref{fig:CAR_all} is most prevalent for `meme' tickers, such as GME, AMC, BB. In stocks that have a broad following outside of the WallStreetBets forum, such as AAPL or MSFT, the CAR appears flat before and after submissions.  
\section{Is information on WSB predictive of returns?}
\label{sec:results_returns}


\paragraph{Initial Exploration}

\begin{table}[ht]
\caption{\textbf{Distribution of Log-Returns}; we present some summary statistics for the next day log-returns for a portfolio invested according to the sentiment of all submissions on WSB (All Submissions), flaired DD posts (Flaired DD) and hand-labeled DD posts (Labeled DD). We compare this to the daily log-returns of all stocks mentioned on WSB (Stock Returns), a randomly selected sample of stocks returns (Random) and the log-returns on the day before submissions are made (Previous). The table presents several summary statistics, as well as the p-value for the mean of the distribution of the daily log-returns to be equal to zero (null hypothesis).} 
\label{tab:dataset_stats}
\centering
{\renewcommand{\arraystretch}{1.2}
\begin{tabular}{c|c c c c c c}
 & $\mu$ & $\sigma$ & skew & kurtosis  & p-value & \# posts \\ 
  \hline
\textit{WSB Returns} &&&&&&\\
All Submissions  &  -0.0177 & 0.22 & -0.20 & 11 & 0.00 & 199,104\\
Flaired DD &  0.0074&  0.19 &  1.29 &  26 & 0.02 &  3,629\\
Labeled DD &  0.0054&  0.14 &  3.19 &  49 & 0.08 &  2,117\\
\hline
\textit{Control Portfolio} &&&&&&\\
\textit{Returns} &&&&&&\\
Stock Returns &  0.0000&  0.04 &  4.81 &  1,934 & 0.52 &  10,411,549\\
Previous  &  0.0238 &  0.35 &  0.50 &  4 & 0.00 &  199,104\\
Random  &  0.0002 &  0.05 &  0.70 &  86 & 0.06 &  198,779\\

\end{tabular}}
\end{table}

We look at the returns of several portfolios constructed from WSB data. We consider the average next day log-returns for investing into all WSB posts mentioning a single ticker according to our sentiment classifier, as well as the next day log-returns for investing into the flaired due dilligence (DD) and labeled DD posts. The results are presented in the top half of Table \ref{tab:dataset_stats}. We observe that investing in all submissions on WSB results in consistently loosing money, with an average next-day log-return of $-0.0177$ and a p-value indicating that the returns are statistically significantly different from zero. The flaired DD posts, on the other hand, which are specifically selected by moderators and our hand-labeled posts both appear to have statistically significant, positive returns (a description of DD posts is described in Section \ref{sec:data}). We perform a similar exercise after splitting the data by year and observe that the forum receives negative or nearly zero returns across all years. In the year 2021, when the infamous GameStop short squeeze occurred, the average next-day log-return for WSB posts was around -0.029 perhaps indicating that even though some forum participants made money on the incident, the subsequent hype and low-quality discussions drove a loosing strategy for the forum as a whole.

We construct several `control' portfolios of interest, presented in the bottom half of Table \ref{tab:dataset_stats}. First, we look at the average, daily returns of the stocks discussed on WSB across all time -- the returns appear to be very close to zero, on average, but with very heavy tails (as indicated by the kurtosis) and a positive skew. The heavy tails (high kurtosis measure) indicates that the stocks mentioned within WSB frequently experience returns that are far from the mean -- extreme return values. The high positive skew, on the other hand, implies that the stocks experience more extremely high returns than low returns. This is consistent with earlier observations that WSB users are swayed by large cumulative abnormal returns of stocks, however, as noted in the previous section, the users tend to be trend followers and the forum sentiment peaks when the asset experiences a price reversion.

We consider the returns of assets on the trading day that a submission is made on WSB (or on the day before if a submission is made outside of trading hours). The high, average positive log-returns indicate that the WSB forum is likely reactive to news, and is consistent with the earlier CAR analysis. We also construct a randomly selected portfolio, where we invest in a given stock proportionally to the number of times it is mentioned in WSB in non-neutral posts, however select days at random. The results indicate a very small, positive return. We anticipate this to be driven by the large returns preceding WSB posts, rather than an inherent ability of WSB users to choose lucrative assets. Appendix \ref{app:clusters_returns} considers returns of portfolios built from the clusters within our \textit{Topic Network} and \textit{Submission Network}. 

\subsection{Predictive signals on WSB}

In Section \ref{subsec:car_analysis}, we consider the extent to which WSB posts are reactive versus anticipatory of asset returns. Through a CAR analysis, we observe that posts are largely reactive, with asset returns increasing in the days \textit{preceding} a posts and decreasing after. In this section, we more rigorously explore short-term signals from the WSB forum and their utility for predicting market returns. We first consider whether WSB activity causes returns in assets, through a Granger-causal analysis. We subsequently consider several signals drawn from due-diligence posts and analyze their effectiveness for predicting returns. 

\subsubsection{Granger causal relationships on WSB}
\label{sec:returns}

We look at whether sentiment is useful for forecasting future returns by conducting a Granger causality test \cite{Granger}. A time series A is said to Granger-cause B if it can be shown that the lagged values of A provide statistically significant information about future values of B, even when lagged values of B are also included in the forecasting exercise. In other words, time series A must provide additional information about the future values of time series B, beyond B's auto-regressive terms.

Our goal is to test whether the sentiments expressed about an asset on WSB have a Granger causal relationship with the future returns of that asset. Our result is displayed as the Wald test statistic, which allows us to estimate a p-value --- our confidence that WSB sentiments about an asset are related to the asset's future returns. 

\paragraph{Sentiment Time Series} To get a time-series of sentiment, we firstly need a sentiment score for each company and date. Let $s^{(j)}_{i,t}$ represent the net sentiment of post $j$, which is calculated by taking the (softmax) output of our sentiment classifier, and calculating the difference between the bullish and bearish outputs (hence $-1 \leq s^{(j)}_{i,t} \leq 1$). We can then say the net sentiment of each stock $i$ on day $t$, denoted $S_{i,t}$, is the sum of $s^{(j)}_{i,t}$ for all posts for that company and day. 
%
$S_{i,t}$, however, has a tendency to increase over time, since the forum has grown in popularity, and most posts are bullish. To resolve this, we normalise $S_{i,t}$ by dividing it by the average number of posts on the forum over the past seven days ($n_t$) -- we denote the normalised sentiment $\hat{S}_{i,t}$. We then calculate the difference in normalised sentiment to get a value for our sentiment time series $\Delta\hat{S}_{i,t}$ where:
\begin{equation}
    \label{eq:sentiment_car}
    \Delta\hat{S}_{i,t} = \hat{S}_{i,t} - \hat{S}_{i,t-1} = \frac{S_{i,t}}{n_t} - \frac{S_{i,t-1}}{n_{t-1}}.
\end{equation}

\begin{table}[h]
\caption{Results of Granger causality tests for log-return for the top 22 most popular stocks on WSB. Main numbers show the Wald test statistic, with $p$ values in parentheses. $^{*}$ are used to indicate significance levels.}
\centering
\begin{tabular}{llllll}
\toprule
 & Obs & 1 & 2 & 5 & 10 \\
\midrule
AAPL & 1338 & 0.140 & 0.047 & 0.439 & 1.177 \\
 &  & $(0.708)$ & $(0.954)$ & $(0.821)$ & $(0.302)$ \\
AMC & 1184 & 9.833 & 36.189 & 21.110 & 10.551 \\
 &  & $(0.002)^{**}$ & $(0.000)^{***}$ & $(0.000)^{***}$ & $(0.000)^{***}$ \\
AMD & 1336 & 0.801 & 2.182 & 1.446 & 1.147 \\
 &  & $(0.371)$ & $(0.113)$ & $(0.205)$ & $(0.323)$ \\
AMZN & 1334 & 0.224 & 0.183 & 1.735 & 2.033 \\
 &  & $(0.636)$ & $(0.833)$ & $(0.123)$ & $(0.027)^{*}$ \\
BABA & 1339 & 1.055 & 1.329 & 1.260 & 0.887 \\
 &  & $(0.305)$ & $(0.265)$ & $(0.279)$ & $(0.545)$ \\
BB & 1212 & 31.792 & 14.402 & 16.236 & 12.083 \\
 &  & $(0.000)^{***}$ & $(0.000)^{***}$ & $(0.000)^{***}$ & $(0.000)^{***}$ \\
DIS & 1326 & 0.410 & 0.760 & 0.731 & 0.950 \\
 &  & $(0.522)$ & $(0.468)$ & $(0.6)$ & $(0.486)$ \\
FB & 1338 & 0.303 & 0.147 & 0.620 & 0.815 \\
 &  & $(0.582)$ & $(0.863)$ & $(0.685)$ & $(0.614)$ \\
GE & 1301 & 0.329 & 0.901 & 1.118 & 1.037 \\
 &  & $(0.566)$ & $(0.406)$ & $(0.349)$ & $(0.409)$ \\
GME & 1284 & 28.820 & 16.922 & 16.206 & 16.749 \\
 &  & $(0.000)^{***}$ & $(0.000)^{***}$ & $(0.000)^{***}$ & $(0.000)^{***}$ \\
MSFT & 1325 & 2.313 & 1.176 & 0.928 & 0.816 \\
 &  & $(0.129)$ & $(0.309)$ & $(0.462)$ & $(0.614)$ \\
MU & 1295 & 0.120 & 0.128 & 0.098 & 0.920 \\
 &  & $(0.729)$ & $(0.88)$ & $(0.992)$ & $(0.514)$ \\
MVIS & 261 & 0.306 & 1.658 & 1.322 & 0.632 \\
 &  & $(0.58)$ & $(0.192)$ & $(0.255)$ & $(0.786)$ \\
NIO & 637 & 0.169 & 0.645 & 0.240 & 0.704 \\
 &  & $(0.682)$ & $(0.525)$ & $(0.945)$ & $(0.721)$ \\
NOK & 1319 & 10.168 & 5.884 & 3.476 & 2.003 \\
 &  & $(0.001)^{***}$ & $(0.003)^{**}$ & $(0.004)^{**}$ & $(0.03)^{*}$ \\
NVDA & 1336 & 0.093 & 0.168 & 0.494 & 0.617 \\
 &  & $(0.761)$ & $(0.845)$ & $(0.781)$ & $(0.801)$ \\
PLTR & 130 & 0.202 & 0.839 & 1.088 & 2.065 \\
 &  & $(0.654)$ & $(0.434)$ & $(0.371)$ & $(0.034)^{*}$ \\
RKT & 167 & 0.168 & 0.529 & 0.283 & 0.287 \\
 &  & $(0.682)$ & $(0.59)$ & $(0.922)$ & $(0.983)$ \\
SNAP & 1022 & 13.782 & 6.907 & 2.815 & 1.811 \\
 &  & $(0.000)^{***}$ & $(0.001)^{***}$ & $(0.016)^{*}$ & $(0.055)$ \\
SNDL & 327 & 0.312 & 3.500 & 1.810 & 1.212 \\
 &  & $(0.577)$ & $(0.031)^{*}$ & $(0.111)$ & $(0.283)$ \\
SPCE & 391 & 3.901 & 1.169 & 1.298 & 2.017 \\
 &  & $(0.049)^{*}$ & $(0.312)$ & $(0.264)$ & $(0.031)^{*}$ \\
TSLA & 1334 & 0.293 & 0.369 & 0.627 & 0.393 \\
 &  & $(0.588)$ & $(0.691)$ & $(0.679)$ & $(0.95)$ \\
\bottomrule
\end{tabular}

\label{tab:Granger}
\end{table}

\paragraph{Granger Causality Results}
We check whether our normalised sentiment measure $\Delta\hat{S}_{i,t}$ is Granger causal of daily stock price returns. We assess whether lagged values of WSB sentiment are useful for forecasting future returns by observing the p-value in our Granger causality test in Table \ref{tab:Granger}. WSB sentiments are lagged between one and ten days with stock return. We use time-series for the top 22 most popular stocks on the forum,\footnote{For consistency with our CAR plots, we include GE and DIS, in addition to RKT and PLTR} starting from the earlier of 1 January 2016, or the first time a stock was mentioned on the forum. The removal of pre-2016 data is motivated by the small size of the forum before this point, which can result in volatile changes in our sentiment measure.

For all 22 stocks, our sentiment measure and the daily log-return time series appear stationary, which we check with an augmented Dickey-Fuller test. We run Granger causality tests at lags of 1, 2, 5, and 10 trading days. The results are shown in Table \ref{tab:Granger}. For most stocks, WSB sentiment is not useful for forecasting returns. This is not true, however, for the meme stocks GME, AMC, NOK and BB, but this fits with the conventional wisdom that retail traders on WSB drove up the prices of these shares in early 2021. 

We do get statistically significant results ($0.01 < p < 0.05$) for AMZN at a ten day lag, PLTR at a ten day lag, SNDL at a two day lag, and SPCE at a one and ten day lag. These results are unlikely to be meaningful however -- given we are running a total of 88 tests, we would expect 4 or 5 results in this range. Given that sentiment is not Granger causal of returns at other lags for these stocks, this is most likely to be a spurious result.

The most interesting statistically significant results come from SNAP, where the test statistic is strongly significant at one and two days ($p < 0.001$), statistically significant at five days ($p < 0.05$), and almost significant at ten days ($p=0.055$). This also coheres with the result of Figure \ref{fig:CAR_stocks}. SNAP is not conventionally considered a meme stock, and it was considerably more discussed prior to 2021.

\subsection{Additional results}

In an additional analysis, we attempt to extract long-term and short-term metrics from WSB, which we consider as potential trade signals and test their predictive power. For our short-term analysis, we focus on extracting signals from our hand-annotated sample of due dilligence (DD) posts. We fit 3,744 models predicting returns at time horizons varying from next day to 24-weeks into the future using various due dilligence post properties. Our models provide some indication of the fact that bullish sentiment in DD posts are predictive of returns at various time horizons. Among other variables that we considered, the existence of a URL (indicative of external references outside of the forum) is  more likely to have a positive effect for models looking at time horizons greater than 8 weeks, and especially at time horizons greater than 15 weeks. For our long-term analysis we consider monthly and weekly changes on the WSB forum versus returns, trading volumes and volatility; we observe that changes in normalized sentiment in an asset in a given month are negatively predictive of returns in the following month. Our full results are presented in Appendix \ref{app:trade_signals}. Overall, our findings suggest that WSB metrics do not have strong, predictive power for market movements, but can be used in conjunction with other factors as potential weak signals in portfolio construction. 
\section{Conclusion}
\label{sec:discussion}

What went on under the hood of the most infamous investor forum, WallStreetBets? This paper takes a data-driven look. We break down the discussion in several different ways: through topic modeling, by sentiment analysis and through a network perspective, used to map the relationships between assets. Our network approach allows us to extract groups or `clusters' of assets which possess similar characteristics on the WSB forum. We conclude that returns for assets within large clusters (those containing many assets) are generally predicted poorly by the forum. However, niche groups of investors may have potential insights into the markets, as indicated by positive returns to WSB submissions within smaller asset clusters.

Our analysis of cumulative abnormal returns (CAR) indicates that WSB investors are, on average, reactive to market news -- they follow the hype and express bullish sentiments precisely at the time when assets reach their peak and, subsequently, begin a CAR price reversal. The pattern is particularly distinct in `meme' stocks, and less pronounced in stocks with a broad following outside of WSB. A final analysis of the average sentiment expressed within the forum indicates that a Granger-causal relationship exists between changes in average sentiment expressed about an asset at future asset returns in several popular assets on WSB. 

The scope of our study leaves several promising directions for future research. A key area for investigation lies in the appropriate way to analyze a non-stationary dataset, such as WSB, and its predictive value across different time periods. Additionally, this analysis presents several methods for of extracting signal from text and network interactions. As unstructured datasets are becoming more and more common, it may be valuable to investigate and standardize additional methods which can be useful for forecasting.

\newpage
\bibliography{bibliography}

\begin{thebibliography}{34}
\providecommand{\natexlab}[1]{#1}
\providecommand{\url}[1]{\texttt{#1}}
\expandafter\ifx\csname urlstyle\endcsname\relax
  \providecommand{\doi}[1]{doi: #1}\else
  \providecommand{\doi}{doi: \begingroup \urlstyle{rm}\Url}\fi

\bibitem[Agrawal et~al.(2018)Agrawal, Azar, Lo, and Singh]{agrawal2018momentum}
S.~Agrawal, P.~D. Azar, A.~W. Lo, and T.~Singh.
\newblock Momentum, mean-reversion, and social media: Evidence from stocktwits
  and twitter.
\newblock \emph{The Journal of Portfolio Management}, 44\penalty0 (7):\penalty0
  85--95, 2018.

\bibitem[Anand and Pathak(2021)]{anand2021wallstreetbets}
A.~Anand and J.~Pathak.
\newblock Wallstreetbets against wall street: The role of reddit in the
  gamestop short squeeze.
\newblock 2021.

\bibitem[Angelov(2020)]{angelov2020top2vec}
D.~Angelov.
\newblock Top2vec: Distributed representations of topics.
\newblock \emph{arXiv preprint arXiv:2008.09470}, 2020.

\bibitem[Araci(2019)]{araci2019finbert}
D.~Araci.
\newblock Finbert: Financial sentiment analysis with pre-trained language
  models.
\newblock \emph{arXiv preprint arXiv:1908.10063}, 2019.

\bibitem[Azar and Lo(2016)]{azar2016wisdom}
P.~D. Azar and A.~W. Lo.
\newblock The wisdom of twitter crowds: Predicting stock market reactions to
  fomc meetings via twitter feeds.
\newblock \emph{The Journal of Portfolio Management}, 42\penalty0 (5):\penalty0
  123--134, 2016.

\bibitem[Baumgartner et~al.(2020)Baumgartner, Zannettou, Keegan, Squire, and
  Blackburn]{baumgartner2020pushshift}
J.~Baumgartner, S.~Zannettou, B.~Keegan, M.~Squire, and J.~Blackburn.
\newblock The pushshift reddit dataset.
\newblock In \emph{Proceedings of the international AAAI conference on web and
  social media}, volume~14, pages 830--839, 2020.

\bibitem[Blei et~al.(2003)Blei, Ng, and Jordan]{blei2003latent}
D.~M. Blei, A.~Y. Ng, and M.~I. Jordan.
\newblock Latent dirichlet allocation.
\newblock \emph{Journal of machine Learning research}, 3\penalty0
  (Jan):\penalty0 993--1022, 2003.

\bibitem[Boylston et~al.(2021)Boylston, Palacios, Tassev, and
  Bruckman]{boylston2021wallstreetbets}
C.~Boylston, B.~Palacios, P.~Tassev, and A.~Bruckman.
\newblock Wallstreetbets: positions or ban.
\newblock \emph{arXiv preprint arXiv:2101.12110}, 2021.

\bibitem[Bradley et~al.(2021)Bradley, Hanousek~Jr, Jame, and
  Xiao]{bradley21wsb}
D.~Bradley, J.~Hanousek~Jr, R.~Jame, and Z.~Xiao.
\newblock Place your bets? the market consequences of investment advice on
  reddit's wallstreetbets.
\newblock \emph{The Market Consequences of Investment Advice on Reddit's
  Wallstreetbets (March 15, 2021)}, 2021.

\bibitem[Buz and de~Melo(2021)]{Buz2021wsb}
T.~Buz and G.~de~Melo.
\newblock Should you take investment advice from wallstreetbets? a data-driven
  approach.
\newblock \emph{arXiv preprint arXiv:2105.02728}, 2021.

\bibitem[Chacon et~al.(2022)Chacon, Morillon, and Wang]{chacon2022will}
R.~G. Chacon, T.~G. Morillon, and R.~Wang.
\newblock Will the reddit rebellion take you to the moon? evidence from
  wallstreetbets.
\newblock \emph{Financial Markets and Portfolio Management}, pages 1--25, 2022.

\bibitem[Costola et~al.(2021)Costola, Iacopini, and
  Santagiustina]{COSTOLA2021110021}
M.~Costola, M.~Iacopini, and C.~R. Santagiustina.
\newblock On the “mementum” of meme stocks.
\newblock \emph{Economics Letters}, 207:\penalty0 110021, 2021.
\newblock ISSN 0165-1765.
\newblock \doi{https://doi.org/10.1016/j.econlet.2021.110021}.
\newblock URL
  \url{https://www.sciencedirect.com/science/article/pii/S0165176521002986}.

\bibitem[Devlin et~al.(2018)Devlin, Chang, Lee, and Toutanova]{devlin2018bert}
J.~Devlin, M.-W. Chang, K.~Lee, and K.~Toutanova.
\newblock Bert: Pre-training of deep bidirectional transformers for language
  understanding.
\newblock \emph{arXiv preprint arXiv:1810.04805}, 2018.

\bibitem[Fama and French(2004)]{capm}
E.~F. Fama and K.~R. French.
\newblock The capital asset pricing model: Theory and evidence.
\newblock \emph{Journal of Economic Perspectives}, 18\penalty0 (3):\penalty0
  25--46, September 2004.
\newblock \doi{10.1257/0895330042162430}.
\newblock URL
  \url{https://www.aeaweb.org/articles?id=10.1257/0895330042162430}.

\bibitem[Gao et~al.(2021)Gao, Zhang, Shi, Xu, Zhang, and Zhu]{gao2021review}
R.~Gao, Z.~Zhang, Z.~Shi, D.~Xu, W.~Zhang, and D.~Zhu.
\newblock A review of natural language processing for financial technology.
\newblock In \emph{International Symposium on Artificial Intelligence and
  Robotics 2021}, volume 11884, pages 262--277. SPIE, 2021.

\bibitem[Gentzkow et~al.(2019)Gentzkow, Kelly, and Taddy]{gentzkow2019text}
M.~Gentzkow, B.~Kelly, and M.~Taddy.
\newblock Text as data.
\newblock \emph{Journal of Economic Literature}, 57\penalty0 (3):\penalty0
  535--74, September 2019.

\bibitem[Granger(1969)]{Granger}
C.~W.~J. Granger.
\newblock Investigating causal relations by econometric models and
  cross-spectral methods.
\newblock \emph{Econometrica}, 37\penalty0 (3):\penalty0 424--438, 1969.
\newblock ISSN 00129682, 14680262.
\newblock URL \url{http://www.jstor.org/stable/1912791}.

\bibitem[Hutto and Gilbert(2014)]{Hutto_Gilbert_2014}
C.~Hutto and E.~Gilbert.
\newblock Vader: A parsimonious rule-based model for sentiment analysis of
  social media text.
\newblock \emph{Proceedings of the International AAAI Conference on Web and
  Social Media}, 8\penalty0 (1):\penalty0 216--225, May 2014.
\newblock URL \url{https://ojs.aaai.org/index.php/ICWSM/article/view/14550}.

\bibitem[Monti et~al.(2019)Monti, Frasca, Eynard, Mannion, and
  Bronstein]{Monti2019}
F.~Monti, F.~Frasca, D.~Eynard, D.~Mannion, and M.~M. Bronstein.
\newblock Fake news detection on social media using geometric deep learning.
\newblock \emph{arXiv preprint arXiv:1902.06673}, 2019.

\bibitem[Pagolu et~al.(2016)Pagolu, Reddy, Panda, and Majhi]{PagoluTwitterSent}
V.~S. Pagolu, K.~N. Reddy, G.~Panda, and B.~Majhi.
\newblock Sentiment analysis of twitter data for predicting stock market
  movements.
\newblock In \emph{2016 International Conference on Signal Processing,
  Communication, Power and Embedded System (SCOPES)}, pages 1345--1350, 2016.
\newblock \doi{10.1109/SCOPES.2016.7955659}.

\bibitem[Poterba and Summers(1988)]{mean_reversion}
J.~M. Poterba and L.~H. Summers.
\newblock Mean reversion in stock prices: Evidence and implications.
\newblock \emph{Journal of Financial Economics}, 22\penalty0 (1):\penalty0
  27--59, 1988.
\newblock ISSN 0304-405X.
\newblock \doi{https://doi.org/10.1016/0304-405X(88)90021-9}.
\newblock URL
  \url{https://www.sciencedirect.com/science/article/pii/0304405X88900219}.

\bibitem[Rao et~al.(2012)Rao, Srivastava, et~al.]{TwitterSentiment1}
T.~Rao, S.~Srivastava, et~al.
\newblock Analyzing stock market movements using twitter sentiment analysis.
\newblock 2012.

\bibitem[Rosenfeld et~al.(2020)Rosenfeld, Szanto, and
  Parkes]{rosenfeld2020kernel}
N.~Rosenfeld, A.~Szanto, and D.~C. Parkes.
\newblock A kernel of truth: Determining rumor veracity on twitter by diffusion
  pattern alone.
\newblock In \emph{Proceedings of The Web Conference 2020}, pages 1018--1028,
  2020.

\bibitem[Schou et~al.(2022{\natexlab{a}})Schou, Bucher, Waldkirch, and
  Gr\"{u}nwald]{Schou22}
P.~K. Schou, E.~Bucher, M.~Waldkirch, and E.~Gr\"{u}nwald.
\newblock We did start the fire: r/wallstreetbets, ‘flash movements’ and
  the gamestop short-squeeze.
\newblock \emph{Academy of Management Proceedings}, 2022\penalty0 (1):\penalty0
  14028, 2022{\natexlab{a}}.
\newblock \doi{10.5465/AMBPP.2022.32}.
\newblock URL \url{https://doi.org/10.5465/AMBPP.2022.32}.

\bibitem[Schou et~al.(2022{\natexlab{b}})Schou, Bucher, Waldkirch, and
  Gr{\"u}nwald]{schou2022we}
P.~K. Schou, E.~Bucher, M.~Waldkirch, and E.~Gr{\"u}nwald.
\newblock We did start the fire: r/wallstreetbets,‘flash movements’ and the
  gamestop short-squeeze.
\newblock In \emph{Academy of Management Proceedings}, volume 2022, page 14028.
  Academy of Management Briarcliff Manor, NY 10510, 2022{\natexlab{b}}.

\bibitem[Semenova and Winkler(2021)]{semenova2021reddit}
V.~Semenova and J.~Winkler.
\newblock Reddit's self-organised bull runs: Social contagion and asset prices.
\newblock \emph{arXiv preprint arXiv:2104.01847}, 2021.

\bibitem[Sul et~al.(2017)Sul, Dennis, and Yuan]{TwitterSentiment2}
H.~K. Sul, A.~R. Dennis, and L.~I. Yuan.
\newblock Trading on twitter: Using social media sentiment to predict stock
  returns.
\newblock \emph{Decision Sciences}, 48\penalty0 (3):\penalty0 454--488, 2017.
\newblock \doi{https://doi.org/10.1111/deci.12229}.
\newblock URL \url{https://onlinelibrary.wiley.com/doi/abs/10.1111/deci.12229}.

\bibitem[Tan and Tas(2021)]{Duz2021}
S.~D. Tan and O.~Tas.
\newblock Social media sentiment in international stock returns and trading
  activity.
\newblock \emph{Journal of Behavioral Finance}, 22\penalty0 (2):\penalty0
  221--234, 2021.
\newblock \doi{10.1080/15427560.2020.1772261}.
\newblock URL \url{https://doi.org/10.1080/15427560.2020.1772261}.

\bibitem[Traag et~al.(2019)Traag, Waltman, and Van~Eck]{traag2019louvain}
V.~A. Traag, L.~Waltman, and N.~J. Van~Eck.
\newblock From louvain to leiden: guaranteeing well-connected communities.
\newblock \emph{Scientific reports}, 9\penalty0 (1):\penalty0 5233, 2019.

\bibitem[Vosoughi et~al.(2018)Vosoughi, Roy, and Aral]{Vosoughi2018}
S.~Vosoughi, D.~Roy, and S.~Aral.
\newblock The spread of true and false news online.
\newblock \emph{Science (New York, NY)}, 359\penalty0 (6380):\penalty0
  1146--1151, 2018.
\newblock ISSN 1095-9203.

\bibitem[Wan et~al.(2021)Wan, Yang, Marinov, Calliess, Zohren, and
  Dong]{Wan2021}
X.~Wan, J.~Yang, S.~Marinov, J.-P. Calliess, S.~Zohren, and X.~Dong.
\newblock Sentiment correlation in financial news networks and associated
  market movements.
\newblock \emph{Scientific Reports}, 11\penalty0 (1):\penalty0 3062, 2021.
\newblock ISSN 2045-2322.
\newblock \doi{10.1038/s41598-021-82338-6}.
\newblock URL \url{https://doi.org/10.1038/s41598-021-82338-6}.

\bibitem[Wang and Luo(2021)]{wang2021predicting}
C.~Wang and B.~Luo.
\newblock Predicting \$ gme stock price movement using sentiment from reddit
  r/wallstreetbets.
\newblock In \emph{Proceedings of the Third Workshop on Financial Technology
  and Natural Language Processing}, pages 22--30, 2021.

\bibitem[Winkler and Semenova(2021)]{Semenova21}
J.~Winkler and V.~Semenova.
\newblock {Reddit's self-organised bull runs: Social contagion and asset
  prices}.
\newblock INET Oxford Working Papers 2021-04, Institute for New Economic
  Thinking at the Oxford Martin School, University of Oxford, Jan. 2021.
\newblock URL \url{https://ideas.repec.org/p/amz/wpaper/2021-04.html}.

\bibitem[Witts et~al.(2021)Witts, Tortosa-Ausina, Arribas,
  et~al.]{witts2021irrational}
D.~W. Witts, E.~Tortosa-Ausina, I.~Arribas, et~al.
\newblock The irrational market: Considering the effect of the online community
  wall street bets on financial market variables.
\newblock Technical report, 2021.

\end{thebibliography}

\newpage
\appendix
\section*{Appendix}

\section{Extended Literature Review}
\label{app:lit_review}
   The WSB forum received limited academic interest until the GME short squeeze in early 2021 made WSB widely familiar. Most papers written on the forum have focused on the GME short-squeeze (and sometimes other `meme' stocks) -- either looking quantitatively at the effect WSB had on the GME stock price, or more qualitatively, trying to understand the emergence of WSB as a social phenomenon \cite{Schou22, anand2021wallstreetbets, wang2021predicting, COSTOLA2021110021}. Few papers have looked at the forum more broadly, one exception is \cite{boylston2021wallstreetbets}, which provided a qualitiative assessment of how the WSB community has developed. Winkler and Semenova \cite{Semenova21} were some of the first authors to look at the behaviour of WSB members in relation to retail trading, analysing the effect of sentiment contagion on the forum and its effect on increasing trading volume. 
    
    More recently, we have started to see papers trying to assess if WSB posts can be used to predict stock price changes. One of the first papers on this subject looked at the returns of the top 20 stocks mentioned on WSB, and found that an equally weighted portfolio of these stocks significantly outperformed the S\&P 500 \cite{Buz2021wsb}. This finding should be treated with caution, as the authors based the portfolio on the most popular stocks from 2018-2021, and used the same time period to measure performance, meaning that the forum may simply prefer stocks that have performed well (there is also no accounting of volatility). The authors also identified `buy' signals from the forum by doing a word count of `buy' and `sell' across posts, and found that a `buy' signal from the forum preceded stock price increases that were higher than the S\&P 500, and also higher than days where no buy signal was found. However, the authors again appeared not to risk-adjust the returns in any way, and found significantly less impressive returns prior to 2021. Later work from other researchers has found WSB sentiment impacts volume and volatility, but it is not a useful price signal \cite{witts2021irrational, chacon2022will}.
    
    The other major paper looking at WSB and returns takes a different approach: focusing on posts marked as `due diligence' (DD) \cite{bradley21wsb}. Such posts are typically more analytical than other WSB posts, and are used by posters to inform others of their trading positions and the reasons for them. The authors found that from 2018-2020, DD recommendations are significant predictors of one-month ahead returns, and comments are incrementally useful when they agree with the sentiment expressed by the post author. However, from 2021 on, the informativeness of DD posts disappears, which the authors infer as evidence that the influx of users after the Gamestop short squeeze have decreased the quality of advice on the platform.

\section{Detailed Data Appendix}
\label{app:data}

\subsection{Data scraping}
\label{subsec:data_scraping}
Submission and comment data from WSB is collected through the Pushshift Reddit database \cite{baumgartner2020pushshift}, which provides an archive of all posts and comments on Reddit. We collect data from the inception of the forum in April 2012 up to 24 June 2022. Posts on WSB typically refer to companies by their stock symbol, and we use the work of \cite{Semenova21} to obtain a list of stock symbols (from Yahoo Finance and Compustat) and extract these from the submission text (with upper-case matching). For symbols that form common words/acronyms (e.g. \textit{\$WISH}), we only match a post with that particular ticker if the symbol is preceded by a dollar sign \$. We inspected several thousand posts to ignore symbols referring to companies that are virtually never discussed on the forum, but whose symbol is used for other references. For example, the appearance of the letters SI almost always refers to short-interest, and never to the associated company: Silvergate Capital Corp. We also ignore any reference to SPY (S\&P 500), since we want to use this for baseline.

For simplicity, we only keep companies traded on the NYSE or NASDAQ exchanges, since this allows us to operate in a consistent timezone. This is also a fairly innocuous step, since the vast majority of WSB activity is focused on the US and companies that trade in US markets. After all these steps are taken, we filter our data to only include posts that mention exactly one stock.

\begin{figure}[thpb]
\centering
\includegraphics[width=10cm]{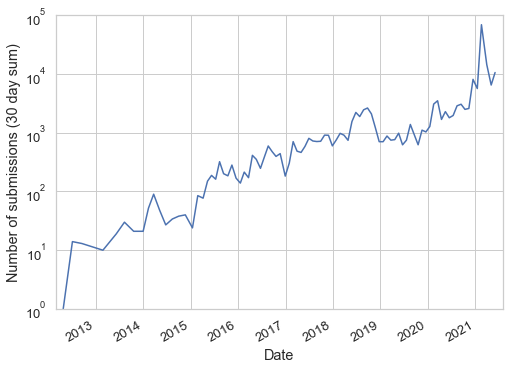}
\caption{Total number of new posts on WSB every 30 days}
\label{fig:submissions_count}
\end{figure}

Stock price data comes from \href{https://www.alphavantage.co/}{AlphaVantage}. AlphaVantage helpfully provides an adjusted close price to take into account events like dividends and stock splits, which we use to calculate returns. We collect a daily time series for each stock in our data set, and we remove posts when AlphaVantage did not have data on a particular stock. At the end of this process, we ended up with 202,681 submissions, with 6,320,968 corresponding comments. As forum activity has increased by several orders of magnitude since inception (Figure \ref{fig:submissions_count}), our data is concentrated in more recent months, with the 2021 data accounting for over 60 percent of posts, despite only accounting for around five percent of the forum's lifetime.

\subsection{Due Diligence (DD) posts}
\label{subsec:data_dd_posts}

To obtain the subset of due diligence (DD) reports in the forum, we firstly filter to posts that have the DD flair. Flairs are used on subreddits to categorise posts, and posts that are flaired as DD are supposed to be moderated by WSB admins. This is the same method used by \cite{bradley21wsb}. We note, however, two issues with this approach: firstly, the DD flair does not appear to be used in any significant way until July 2018, so this excludes any older posts. Secondly, we noticed upon reading the DD posts that quality control is limited in our dataset. The \href{https://www.reddit.com/r/wallstreetbets/wiki/linkflair/}{flair guidelines} state that DD posts must be a high-effort, longer, researched posts, but many posts in the data set fail to meet this criteria. This is partially because Pushshift retains many posts that are later removed by moderators -- these posts are straightforward to remove by looking up each DD submission on the Reddit API -- but many posts that contain the DD flair still do not appear to meet the criteria. For example:

\begin{quote}
   \textit{ Very short DD , but potentially bigly (not investment advice) will be going in today myself on this. NOK [Nokia] literally going to the moooooonnnnnnn}
\end{quote}

To resolve these issues, we take two additional filtering steps beyond \cite{bradley21wsb}. Firstly, we supplement the posts containing the DD flair with posts that contain the DD acronym. Secondly, we manually review every single potential DD post (i.e. flaired posts that are not removed by moderators, or contain the text `DD'), and we remove any post that does not appear to be a valid DD. For us to consider a post a valid DD, we require: 
\begin{enumerate}
    \item The post is at least 20 words long 
    \item The post is primarily about one company
    \item The post has a positive or negative sentiment on the company's future price
    \item The post attempts to provide a reason for the future price to change
\end{enumerate}

Inevitably, there is some subjectivity in this filtering, especially for the third criteria. We attempt to be quite permissive in assessing these criteria; for example, we would allow news about a new product release to be used as a reason (even though this is very likely to already be incorporated into the price). We do not, however, permit reasons that are excessively vague; for example, saying a company's price will increase because `fundamentals are good' is insufficient without some added specificity of what is good about the fundamentals. These filtering criteria remove around 40 percent of our candidate posts, leaving us with 3,650 DD posts.

The DD posts are sufficiently small in number that we can manually label each post as bullish/positive or bearish/negative manually (neutral posts are removed). We find 77 percent of posts are bullish, and 23 percent are bearish. 

\subsection{Sentiment detection}
\label{subsec:data_sentiment_detection}

For the forum as a whole there are far too many posts for manual classification. Previous papers have either relied on counting keywords (for example, `buy', `sell', `call', `put' etc. \cite{Buz2021wsb, bradley21wsb}), or they have used off-the-shelf rule based sentiment classifiers \cite{wang2021predicting}, such as VADER \cite{Hutto_Gilbert_2014}.

The advantage of such approaches is their simplicity, but they have several problems. Choosing keywords/phrases is inherently arbitrary -- it could easily miss posts that are clearly positive/negative because they don't contain the required keywords, and it is vulnerable to mistakes when authors negate phrases in the dictionary (e.g. `calls are going to be worthless'). Algorithms like VADER, by contrast, lack the specific domain vocabulary of the WSB forum, hence their performance can be quite poor \cite{wang2021predicting}.

Another issue with using VADER is that we are not interested in the general sentiment of WSB posts \textit{per se}, instead, we are interested in whether the poster has a bullish, bearish, or neutral attitude towards the future price of an asset. To clarify, suppose a trader talks about how much money they made from a recent trade on Tesla; the post is likely to contain positive sentiment towards Tesla, but it is backwards looking, and the author may take no position on the future value of Tesla shares. We want to categorise such a post as neutral, despite it having an overall positive valence.

To resolve these issues, we firstly hand-code a random sample of 4,000 WSB posts, categorising them as bullish, bearish, or neutral depending on what the author's attitude is on future price increases. Of the posts in the sample 41 percent are bullish, 37 percent are neutral and 22 percent are bearish. \footnote{Since bearish posts tend to be a minority on the forum, bearish posts were up-sampled to reduce class imbalance.} We then take a pre-trained BERT \cite{devlin2018bert} model called FinBERT \cite{araci2019finbert}, which we fine-tune using our manually labelled sample. FinBERT has been trained to perform sentiment detection on financial news, which is helpful as it can pick up on financial jargon used in the forum.

The sample was split, with 75 percent of data used for training, 10 percent for validation and 15 percent for testing. The maximum text length was 256 words (shorter posts were padded, longer posts were truncated), and we ran batch sizes of 8\footnote{Fine tuning was performed on a Nvidia RTX 3070 GPU, which ran into memory issues when larger batch sizes were attempted}. Validation accuracy was assessed every 100 batches, and training was stopped when cross-validation accuracy failed to increase after 10 evaluations.

\begin{table}[h]
\caption{Confusion matrix showing the results of the fine-tuned FinBERT classifier on the test data ($n=604$)}
\label{table_example}
\begin{center}
\begin{tabular}{lrrrr}
\toprule
& & & \textbf{Prediction} & \\
& & Bullish & Bearish & Neutral \\
\midrule
& Bullish & 173 (28.6\%) & 30 (5.0\%) & 49 (8.1\%) \\
\textbf{Label} & Bearish & 14 (2.3\%) & 95 (15.7\%) & 21 (3.5\%) \\
& Neutral & 42 (7.0\%) & 30 (5.0\%) & 150 (24.8\%) \\
\bottomrule
\end{tabular}

\end{center}
\label{tab:confusionmatrix}
\end{table}

Our final model had a test accuracy of 69.2 percent, which we consider sufficient due to the challenging task of predicting future-facing sentiment. When a sample of the manually annotated data was checked by a second human, we found labelling disagreement in around 10 percent of cases. Around half the time this was due to human error, but in around 5 percent of cases, it was genuinely ambiguous. Furthermore, the performance is substantially better than a majority class baseline (40.7 percent). The confusion matrix is shown in Table \ref{tab:confusionmatrix}. 

\newpage
\section{Topic Model}
\label{app:topic_model}
We present a more detailed description of the tickers present in different topics from our topic model in Table \ref{tab:appendix_topic_clusters}.
For each cluster, we report the top most mentioned tickers, by number of posts associated with it. We assign up to 10 tickers per cluster for readability, but we also report the number of posts in total accross all tickers as well as the total number of tickers in that cluster.

\begin{table}[htbp]
  \centering
  \caption{\textbf{Sizes and contents of topic clusters}}
  \label{tab:appendix_topic_clusters}
  {\renewcommand{\arraystretch}{0.3}}%
    \begin{tabular}{|rp{10.25em}rr|}
    \toprule
    \multicolumn{1}{|l|}{Cluster} & \multicolumn{1}{l|}{Tickers} & \multicolumn{1}{l|}{No. of Posts} & \multicolumn{1}{l|}{No. of Companies in Cluster}  \\
    \midrule
    0     & AMZN, SNDL, HMNY, MVIS, WISH, CLNE, CCL, DKNG, SUNE, CTRM & 12,373 & 41  \\\midrule
    1     & {AAPL} & 2,530  & 1  \\ \midrule
    2     & {MSTX, AVXL, MNKD, OPK, AMDA, AUPH, SENS, AMRN, ADMP, CARA}  & 3,312  & 44  \\ \midrule
    3     & {BB, BBBY, BNGO, BRK, BBY, DB, BLNK, ABNB, BBW, BTT} & 9,898  & 12  \\ \midrule
    4     & {MU, TLRY, ACB, CGC, APHA, CRON, AMAT, MO, OGI, HEXO} & 6,557  & 17  \\ \midrule
    5     & {NFLX, SQ, ATVI, PLUG, WMT, SBUX, LULU, CRM, CMG, MCD} & 10,045 & 72  \\ \midrule
    6     & {FB} & 1,580  & 1  \\ \midrule
    8     & {SNAP, NVDA, NAKD, NAK, FSLY, TTD, DBX, KR, TWLO, GNC} & 5,763  & 35  \\ \midrule
    9     & {AMC} & 19686 & 1  \\ \midrule
    11    & {NOK, RKT, WKHS, SOS, ZOM, QS, SRNE, TRCH, ASO, LFIN} & 11,349 & 33  \\ \midrule
    13    & {MSFT, MS} & 1,849  & 2  \\ \midrule
    14    & {NIO, XPEV} & 2,160  & 2  \\ \midrule
    15    & {INO, OCGN, PFE, NVAX, MRNA, CVS, JNJ, AZN, VXRT, TEVA} & 2,474  & 13  \\ \midrule
    16    & {BABA} & 1,175  & 1  \\ \midrule
    18    & {USO, UVXY, CHK, WTI, XOM, SVXY, NAT, RIG, UCO, BP} & 1,001  & 12  \\ \midrule
    19    & {SPY, TSLA, AMD, PLTR, SPCE, NKLA, QQQ, PRPL, GM, ROKU} & 28,438 & 30  \\ 
    \bottomrule
    \end{tabular}%
\end{table}%

\section{CAR for Popular Stocks}
\label{app:CAR}
\begin{figure*}[ht!]
\centering
 \includegraphics[width = 0.9\textwidth]{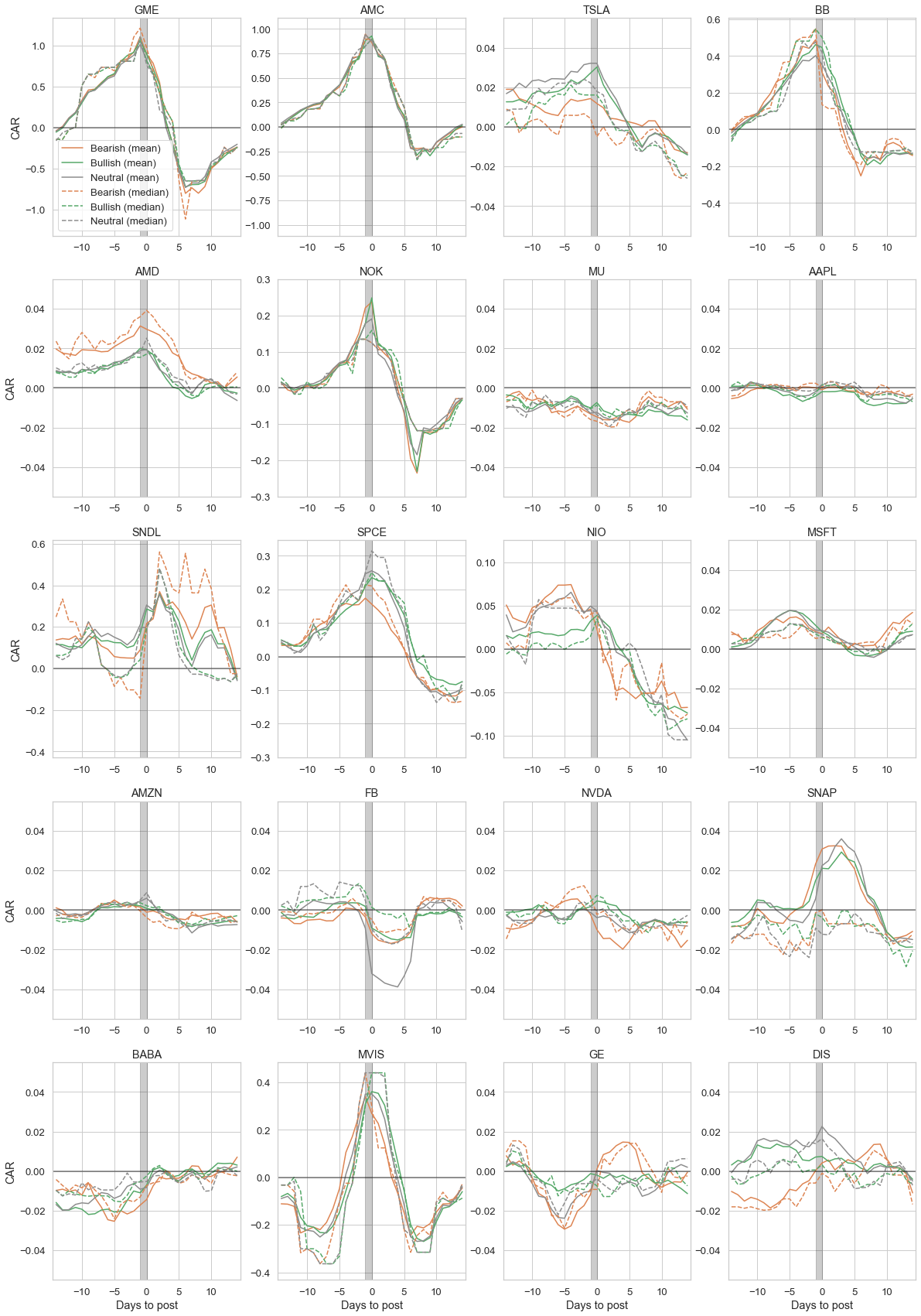}
\caption{Average cumulative abnormal return 14 trading days before and after post submission, grouped by post sentiment for each of the 20 most popular stocks on WSB.}
\label{fig:CAR_stocks}
\end{figure*}

Figure \ref{fig:CAR_stocks} shows a breakdown of the CAR plots for the twenty most popular stocks on the forum. \footnote{Palantir Technologies (PLTR) and Rocket Companies (RKT) are more popular stocks on WSB than General Electric (GE) or Disney (DIS), however, both stocks had their IPOs in 2020, so there is not sufficient data to fit the CAPM model for these stocks.} We see similar patterns to Figure \ref{fig:CAR_main} for smaller `memey' stocks like GME, AMC, BB, SPCE, and NOK. This pattern can also be observed for TSLA and AMD. Other large companies like AAPL, MSFT, AMZN, NVDA, GE and BABA are fairly flat in their pattern, suggesting a weaker relationship between sentiment and returns. The only two cases where sentiment does seem to meaningfully precede peaks in CAR are SNDL and SNAP. The performance of SNDL (a cannabis company) appears by a spike in popularity in late January 2021, which preceded a tripling in the share price over the next few weeks. The performance of SNAP also appears to be driven by outliers, as its median CAR chart is quite flat compared to its mean. Overall, the CAR analysis suggests that besides a few outliers, WSB sentiment appears to be a lagging indicator of stock price changes, not a leading one. Our interactive online dashboard allows one to individually inspect stock CAR plots, as well as returns, leading up to and following posts on WSB.

\section{Discussion of Clusters and Returns}
\label{app:clusters_returns}
\paragraph{Author Clusters}
Figure \ref{fig:sub_network_observed} presents eleven investor clusters, where assets are connected to other assets based on whether a similar subset of authors interacts in discussions about both assets simultaneously. In this section, we investigate the profitability of investors on WSB within these different clusters, as well as the correlation of asset returns within these clusters. We remind the reader of the most central assets within each of the discussion clusters: 0) DAL, UAL, 1) PLTR, 2) AMD, AAPL, MU, 3) SLV, GLD, 4) TSLA, SPY, 5) GME, AMC, 6) CCL, RCL, 7) LMT, BA, 8) JNUG, NUGT, 9) TLRY, CGC, 10) NKLA.

\begin{table}[ht]
\caption{\textbf{Distribution of Log-Returns for Investor Clusters}; we present some summary statistics for the next day log-returns for a portfolio invested according to the sentiment of submissions in different asset clusters on WSB as presented in Figure \ref{fig:sub_network_observed}. The table presents several summary statistics, as well as the p-value for the mean of the distribution to be equal to zero. Cluster \textbf{8} has insufficient observations and, therefore, is not considered.} 
\label{tab:return_clusters}
\centering
{\renewcommand{\arraystretch}{1.2}
\begin{tabular}{c|c c  c c c | c}
   &&&&&&  \textit{Within Cluster}\\ 
 & $\mu$ & $\sigma$ & kurtosis  & p-value & \# posts & \textit{Asset Correlation}\\ 
  \hline
\textit{Cluster} &&&&&&\\
\textbf{0}& -0.0211& 0.09& 1.99& 0.00& 593& 0.80\\
\textbf{1}& -0.0101& 0.09& 1.58& 0.00& 5,567& 0.36\\
\textbf{2}& -0.0016& 0.04& 18.23& 0.00& 14,089& 0.22\\
\textbf{4}& -0.0012& 0.07& 44.03& 0.04& 13,006& 0.23\\
\textbf{5}& -0.0791& 0.44& 1.56& 0.00& 61,787& 0.17\\
\textbf{6}& 0.0062& 0.10& 5.99& 0.21& 398& 0.84\\
\textbf{7}& 0.0077& 0.09& 2.32& 0.00& 2,324& 0.35\\
\textbf{9}& -0.0241& 0.21& 3.81& 0.00& 867& 0.60\\
\textbf{10}& -0.0095& 0.12& 1.34& 0.11& 430& 0.10\\
\end{tabular}}
\end{table}

Table \ref{tab:return_clusters} presents our key result. We observe that portfolios built using opinions from WSB fairly consistently loose money. The tendency is particularly stark in the "meme" stock cluster -- cluster \textbf{5}. This perhaps highlights the tendency for less sophisticated investors to participate in conversations surrounding these assets. A distinct, more profitable cluster appears to be cluster \textbf{7}. Unlike several of the other asset clusters, the assets appear to be more loosely connected, even though a potential unifying theme of large-scale engineering / space exploration is noticeable, since the cluster contains BA (Boeing), LMT (Lockheed Martin) and SPCE (Virgin Galactic Holdings). This analysis, in combination with our portfolio-building exercise with DD posts, highlights that overall investors on WSB appear to be trend-followers and fail to predict the market, however, small pockets of investors may have meaningful insights and outperform the market. 

\paragraph{Topic Clusters}

\begin{table}[htbp!]
\caption{\textbf{Distribution of Log-Returns for Topic Clusters}; we present some summary statistics for the next day log-returns for a portfolio invested according to the sentiment of submissions in different asset clusters on WSB as presented in Figure \ref{fig:topic_network_observed}. The Table \ref{tab:Log-Returns-Topic-Clusters} below presents several summary statistics, as well as the p-value for the mean of the distribution to be equal to zero. The clusters of GNUS (cluster 7), GTE (cluster 10), MGNI (cluster 12) and RC (cluster 17) are not considered as they do not meet a minimal threshold of 1000 posts. \\} 

\break 
\label{tab:Log-Returns-Topic-Clusters}

{\renewcommand{\arraystretch}{0.8}}
  \centering
  
    \begin{tabular}{c|cccccc|}
          &       &       &       &       &       &  \\
          & \textbf{$\mu$}    & \textbf{$\sigma$} & kurtosis & p-value & \# posts & \textit{\# tickers in cluster} \\
    \midrule
    \textit{cluster} &       &       &       &       &       &  \\
    \textbf{0}     & -0.0135 & 0.14  & 10.60 & 0.00  & 12,373 & 41   \\
    \textbf{1}    & 0.0027 & 0.03  & 4.33  & 0.00  & 2,530  &  1  \\
   \textbf{ 2 }    & -0.0211 & 0.13  & 9.57  & 0.00  & 3,312  & 44   \\
   \textbf{ 3 }    & -0.0301 & 0.18  & 2.71  & 0.00  & 9,898  & 12   \\
   \textbf{ 4 }    & -0.0168 & 0.13  & 21.04 & 0.00  & 6,557  & 17   \\
   \textbf{5}     & -0.0005 & 0.06  & 30.90 & 0.41  & 10,045 & 72   \\
   \textbf{6}     & -0.0018 & 0.03  & 17.60 & 0.03  & 1,580  & 1  \\
    \textbf{8}    & -0.0029 & 0.06  & 58.04 & 0.00  & 5,763  & 35   \\
    \textbf{9}    & -0.0639 & 0.37  & 0.57  & 0.00  & 19,686 & 1  \\
    \textbf{11}    & -0.0742 & 0.16  & 3.63  & 0.00  & 11,349 & 33   \\
    \textbf{13}    & 0.0023 & 0.02  & 4.06  & 0.00  & 1,849  &  2  \\
    \textbf{14}   & 0.0130 & 0.09  & 1.60  & 0.00  & 2,160  &  2  \\
    \textbf{15}    & -0.0095 & 0.13  & 10.90 & 0.00  & 2,474  & 13   \\
    \textbf{16}    & 0.0026 & 0.03  & 2.69  & 0.00  & 1,175  &  1  \\
   \textbf{ 18 }   & -0.0016 & 0.06  & 3.13  & 0.40  & 1,001  & 12   \\
    \textbf{19}    & -0.0009 & 0.07  & 3.41  & 0.03  & 28,438 & 30   \\
    \end{tabular}%
\end{table}%


In Section \ref{subsec:topic_model_data} we introduced our topic model and topic graph: Asset clusters identified by our topic graph isolate either individual tickers with highly distinct topics in their discussions, or broader groups of clusters with less distinct and more widely shared discursive environments; ticker topic clusters are displayed in Table \ref{tab:appendix_topic_clusters}. Table \ref{tab:Log-Returns-Topic-Clusters} assesses the statistical properties and behaviour of these asset clusters. 

We observe, that most of the identified clusters with statistically significant p-values at the 1\% threshold loose money. This suggests that on the whole, as reported above, strategies identified by mining the topic graph loose money. However, similarly to Table \ref{tab:return_clusters}, certain clusters topic clusters exhibit positive daily returns, on average: cluster 1 (AAPL), 13 containing Microsoft and Morgan-Stanley, 14 containing NIO and XPEV two electric carmakers, and cluster 16 containing Alibaba. We notice that the smaller clusters with fewer assets perform better than much larger clusters, as they are the only ones displaying positive returns on average. On one hand, this may be driven by the systematic over-performance of tech stocks such as AMC, APPL and MSFT over the period. However, not all clusters with positive returns identified by the topic graph are composed of tech stocks, and conversely, not all the small clusters have positive returns. This allows us to suggest that tighter, more focused topic environments may indicate the presence of specialized `pockets' of investors with better understanding and insights into the market. 

\section{Short and long term trade signals}
\label{app:trade_signals}
\subsection{Due diligence reports}
\label{app:short_term_dd}

\paragraph{Signal Extraction}

    Since DD reports only form a small fraction of total WSB activity, the methods used for investigating the relationship between average sentiment and future returns are not appropriate, as these relied on averaging over thousands of data points per stock, and constructing time series for the Granger causality tests. Instead, it is better to consider whether the event of each DD post precedes a change in return. We largely follow the methods of \cite{bradley21wsb} in our analysis of the DD reports. 
    
    Our main variable of interest is the sentiment of posts on day $t$ for company $i$, which we will denote $d_{i,t}$. This is very similar to our definition of $S_{i,t}$ from the previous section, however in this case we separate bullish posts from bearish posts and fit models on each of them instead of computing a net score. The justification for this choice is that we wanted to consider the effect of variables beyond sentiment, and this is much more straightforward when the posts are treated separately.
    
    Beyond sentiment, we defined a series of additional variables that might be a relevant indicator of the quality of a post:
    \begin{enumerate}
        \item The number of comments a post receives. We consider the submission to be a comment so each submission has at least one comment.
        \item The number of words the post contains.
        \item The maximum depth of a submission comment. We consider the submission itself to have a depth of zero, comments replying to the submission have a depth of one, comments replying to comments replying to the submission have a depth of two and so on. This value is divided by the log number of comments in the post, else it is highly correlated with our first variable.
        \item A binary indicator whether a post contains a URL. URLs could be to another Reddit post or an external source, but we did not include urls to images.
        \item A binary indicator whether a post was `proactive'. We consider a post to be proactive if the post occurred when the CAR of the stock was not statistically significantly different from zero (using a t-test). Contrast this to `reactive' posts, which are talking about companies whose price is substantially different from what we would expect under the CAPM.
    \end{enumerate}
    
    All variables were aggregated by summing across each unique day/company combination, and the first two variables were also log transformed. Since two of our variables depend on actions taken after the post is submitted, we only include comments that occur within 24 hours of the original post.
    
    The choice of each of these features was motivated by various considerations. Number of comments and maximum depth can both be considered as graph features when the Reddit submission and successive comments are represented as a directed acyclic graph, also known as an information cascade. Previous research on Twitter has found that graph features can be predictive of the content of Tweets, which has been a particularly active area of research in fake news detection \cite{Vosoughi2018}, \cite{rosenfeld2020kernel},\cite{Monti2019}. If Reddit users react differently to posts with more predictive power (perhaps because they are better analysed), then such graph features could be useful for our models.
    
    The other three features serve as proxies for the quality of a post. Longer posts may have more detailed robust analysis, and therefore may be more predictive of price changes. Posts that contain URLs are providing sources, which again could be an indicator of a more reliable signal. The proactive feature is inspired by one of the previous WSB papers \cite{Buz2021wsb}, which suggested that when a submission is created when the price of an asset has not changed recently, then it might be more likely to contain a useful signal compared to posts about assets whose price is already very far from its typical value (and therefore likely to experience mean-reversion).
    
    For each day and company in our dataset, we calculate a net DD sentiment score. We then fit the following fixed effects model:
    
    \begin{equation}
        \begin{gathered}[b]
            r_{i,[t,t+m]} = \beta_1 d_{i,t} + \beta_2 x_{i,t,k} + \beta_3 r_{i,[t,t-1]}  \\ 
            + \, \beta_4 CAR_{i,t} + \gamma_t + \epsilon_{i,t}.
        \end{gathered}
    \end{equation}
    $r_{i,[t,t+m]}$ is the (log) return between day $t$ and day $t + m$, $d_{i,t}$ represents our sentiment variable, and $x_{i,t,k}$ is one of our additional variables, where $k$ denotes which variable is being used. $r_{i,[t-1,t]}$ is the return of the stock when comparing the previous day price to day $t$, and $\gamma_t$ is our fixed time effects respectively. Time fixed effects ensure that we control for overall movements of the markets, but including fixed effects for firms is inappropriate for two reasons. Firstly, it would mean that our model was including future information about a company's performance, and secondly, if the forum is good at picking stocks that generally outperform the market (even if they do not get the timing especially precise), then we want to be able to observe that signal in our model. To estimate our fixed effects and control variables accurately, we include closing price data for all stocks mentioned in the forum at least once, and include prices from 2017 July 17 up to 21 June 2021, which is the date range of our DD sample. This gives each model around 1.5 million data points to estimate our time effects and control variables.
    
    Note that the comparison date $t$ is defined as being the next open or close price of the asset 24 hours after the post was published. This is done to block reverse causality, since it is possible that number of comments or maximum comment depth increase after price changes, which would show as a misleading association in our model.
    
    Having previous day return and CAR included in our model has a couple of benefits. Firstly, depending on the size of $m$, the control variables can help to control for price changes resulting from momentum or mean reversion \cite{mean_reversion}. Second, as we saw in our previous section, forum activity tends to be highest when CAR is very high, so including these controls ensures that any effect measured is not simply a proxy for prices already being high. To ensure our control variables are estimated precisely, our model includes all companies and trading days from the creation date of the earliest DD post, up to the creation date of the last DD post.
    
    We fit 3,744 models, the number resulting from the multiplication of the following configurations:
    \begin{itemize}
        \item Bullish and bearish posts are considered in separate models -- this makes aggregation of the remaining independent variables considerably more straightforward (2 configurations).
        \item Values of $m$ are considered from one to 26 weeks in one week intervals (26 configurations).
        \item Every model has sentiment as an independent variable, and zero or one of our other independent variables (6 configurations).
        \item Whether the model considers all posts, or just flaired posts (2 configurations).
        \item Whether the model considers all stocks, or excludes meme stocks: GME, AMC, BB, and NOK (2 configurations).
        \item Whether the model considers all posts, posts before 2021, and posts from 2021 (3 configurations).
    \end{itemize}
    
    Having multiple configurations allows us to discern whether any effects we observe are robust to alternative configurations. We therefore not only consider whether a parameter is statistically significant in a single model, but whether it is statistically significant under other considerations.

\paragraph{Due diligence and market returns}

    The results of the fixed effects models for the DD posts are shown in Appendix \ref{app:due_dilligence_tables}, Tables \ref{tab:fe_results_bull_4}, \ref{tab:fe_results_bear_4}, \ref{tab:fe_results_bull_12}, and \ref{tab:fe_results_bear_12}. Since we fitted over three thousand models, it is not practical to show all results, so instead presentation is limited to just the models looking at price changes over four weeks and twelve weeks. The results for the models that only include the flaired DD posts are not included, since it was of little practical importance.

    Note that the sign for the sentiment variable is flipped for the bearish models, so it is always the case that a positive coefficient for the sentiment variable implies that prices are increasing in the same direction the post creator predicted, while negative coefficients imply that prices moved in the opposite direction to the prediction. The signs for other variables is not reversed, so positive coefficients imply increasing asset prices in all model specifications, and vice-versa for negative coefficients.
    
\begin{figure}[thpb]
\centering
\includegraphics[width=10cm]{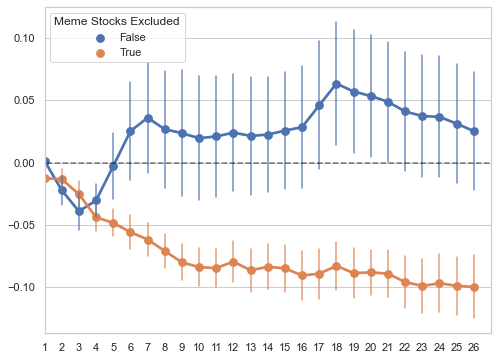}
\caption{The coefficient for the sentiment variable for post-2021 bullish models with no additional features at different time horizons (1-26 weeks). Split by whether meme stocks (GME, AMC, BB, and NOK) are included. Bars indicate 95 percent confidence intervals.}
\label{fig:sentiment_param}
\end{figure}

    Our results indicate that the number of comments and the maximum comment depth are not associated with future price changes. Across all our configurations, number of comments and maximum comment depth were respectively statistically significant in 0.8 and 2.2 percent of the models they were included. The coefficient for sentiment was statistically significant in 19 percent of our models. There are distinct patterns to the model configurations where this occurs. Firstly, the parameter is rarely statistically insignificant in the bearish models. The only time this occurs in the displayed results is in Table \ref{tab:fe_results_bear_4}, for the post-2021 model with meme stocks included and no other independent variables (effect size -0.93, $p<0.05$). Secondly, when sentiment is statistically significant, the coefficient is negative in about three models for every one model where the coefficient is positive. In the bullish four week models shown in Table \ref{tab:fe_results_bull_4}, the sentiment parameter is never positive and statistically significant. Thirdly, in the 543 configurations where sentiment is statistically significant and the coefficient is negative, 63 percent of these models include only the posts published in 2021. This matches a finding from \cite{bradley21wsb}, that asset prices tend to move in the opposite direction of DD predictions made in 2021. Figure \ref{fig:sentiment_param} shows how the sentiment parameter varies when altering the time window of our model. Interestingly, as Figure \ref{fig:sentiment_param} demonstrates, this finding only holds in the models where meme stocks are excluded. If meme stocks are included, returns are negative in the short-run but positive (albeit with high uncertainty) at time horizons longer than 6 weeks. Finally, in the 174 configurations where sentiment is statistically significant and the coefficient is positive, 172 of those configurations contain an additional independent variable. Examples of this phenomenon are apparent in Table \ref{tab:fe_results_bull_12} in some of the models where post word count is included as an additional feature. 

\begin{figure}[thpb]
\centering
\includegraphics[width=10cm]{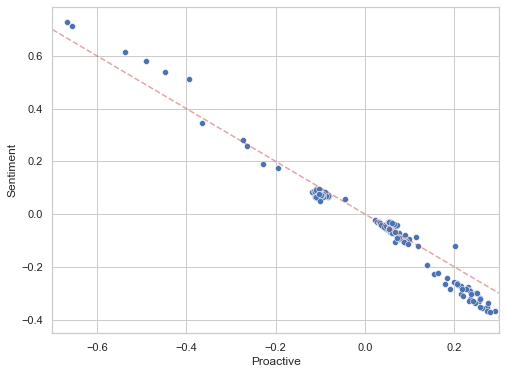}
\caption{Scatterplot showing the parameter values of proactive and sentiment in all models where proactive is statistically significant. The red dashed line shows $y=-x$}
\label{fig:proactive}
\end{figure}

    This leads naturally to a discussion of our remaining independent variables. Post word count is statistically significant in around 21 percent of models where it was included, but when it is statistically significant, the coefficient is negative 97 percent of the time. For example, in Table \ref{tab:fe_results_bull_12}, in the model including all bullish posts, the word count variable has an effect size of -0.0135 ($p < 0.001$). Notice, however, that in the same model, the sentiment parameter is statistically significant with an effect size of 0.0672 ($p < 0.001$), roughly five times the magnitude and the opposite sign. Since the number of comments variable was logged, it turns out that the average post has a logged word count of approximately five, meaning for the typical post these effects washout, and the apparent effects are primarily a result of multicollinearity.
    
    The URL feature is statistically significant in 18 percent of models, and always positive when significant. This feature tends to only be significant in the bearish models, for example, in the pre-2021 bearish model including meme stocks, the URL coefficient is 0.0597 ($p < 0.05$). This effect cannot be explained by multicollinearity, as the sentiment parameter for this model is negative with effect size of -0.0242 (not statistically significant), but since this is a bearish model, that implies that bearish posts with URLs are more likely to precede an asset price increase. The URL parameter is also more likely to show up as positive for models looking at time horizons greater than 8 weeks, and especially at time horizons greater than 15 weeks.
    
    Finally, the proactive variable is statistically significant in 27 percent of models. In bullish models where it is a statistically significant feature, it is always positive, and conversely in bearish models where the proactive variable is statistically significant, it is always negative. While this is promising, the actual effect size of the proactive variable is typically very close in magnitude to the sentiment variable with the opposite sign. For example, in the four-week model with all bullish posts, sentiment has a effect size of -0.0404 ($p < 0.001$), while proactive has an effect size of 0.0443 ($p < 0.001$).Since 80 percent of posts are labelled as proactive under our definition 
    \footnote{While this may seem like a surprisingly high proportion, recall our definition of a 'proactive' post was one with a CAR that was not statistically significantly different from zero, which is going to be true 95 percent of the time when selecting stocks and days at random.}, these effects are largely cancelling out, and once again indicate multicollinearity.

\subsection{Longer-term trade signals}
\label{app:long_term}

\begin{table}[ht!] 
\begin{center}
  \caption{Stock returns versus granular social shocks}
  \label{tab:long-term_signals} 
\begin{tabular}{@{\extracolsep{-2pt}}lccc|ccc} 
\\[-1.8ex]\hline 
\hline \\[-1.8ex] 
 & \multicolumn{3}{c}{Weekly Dependent Variables} & \multicolumn{3}{c}{Monthly Dependent Variables} \\ \\[-1.8ex]
& $r_{i,t+1}$ & $\sigma_{i,t+1}$ & $v_{i,t+1}$ & $r_{i,t+1}$ & $\sigma_{i,t+1}$ & $v_{i,t+1}$ \\ \\[-1.8ex] 
 & (1) & (2) & (3) & (4) & (5) & (6) \\ 
\hline \\[-1.8ex] \\[-1.8ex]
WSB Variables &&&       &&&\\
$M_{i,t}$ & -0.002* & -0.001***& -0.020***       & 0.014 & -0.001*** & -0.026***\\
 & (0.001)& (0.000) & (0.005)     & (0.017) & (0.000) & (0.005) \\
 \\[-1.8ex]
 $\Delta S_{i,t}^N$ & -0.000 & -0.001 & 0.007   & 0.030 & -0.001** & 0.022**\\
 & (0.002)& (0.001) & (0.008)      & (0.036) & (0.000) & (0.009)\\
 \\[-1.8ex]
$S_{i,t}^N$ & 0.004 & 0.001* & 0.015       & -0.103** &  0.002*** & -0.018\\
 & (0.003) & (0.001) & 0.010       & 0.047 & (0.001) & (0.012)\\
\hline \\[-1.8ex]
Controls &&&       &&&\\
$r_{i,t}$ & -0.042*** & 0.022*** & -0.030      & -0.026*** & -0.006*** & 0.045**\\
          & (0.008) & (0.002) & (0.032)      & (0.009) & (0.001) & (0.023)\\
$\sigma_{i,t}$ & -0.138*** & 0.387*** &  -3.355***     & -2.002 & 0.435*** & -4.693***\\
          & (0.022) & (0.006) & (0.090)      & (0.548) & (0.007) & (0.143)\\
\\[-1.8ex]
\hline \\[-1.8ex]
Observations & 13,119 & 13,119 & 13,119 &          9,528 & 9,528 & 9.528 \\ 
R$^{2}_{adj}$ & 0.007 & 0.234 & 0.105 &   0.003 & 0.317 & 0.114\\ \\[-1.8ex] 
\hline 
\hline
\end{tabular} 
\end{center}
\footnotesize{ \textit{Notes}: this table presents the relationship between longer-term signals from WSB and the markets. Columns 1-3 examine the relationship on a weekly timescale, while columns 4-6 examine the relationship on a monthly timescale. \\
*** Significant at 1\% level
** Significant at 5\% level
* Significant at 10\% level} \\ 
\end{table} 

In addition to short-term signals, which look at the sentiments on the forum on a given day, we propose that longer-term trends within the forum may be predictive of market returns. Specifically, we look at the relationship between several signals on the forum on a weekly and monthly timescale, and their relation to returns, volatility and volumes (also at weekly and monthly time periods).

\textit{Momentum:} We propose a metric for momentum on the forum:
\begin{align}
    M_{i,t} = log \left(\frac{n_{i,t}/n_t}{n_{i,t-1}/n_{t-1}}\right)
\end{align}
which captures the log of the fraction of posts discussing asset $i$ in week $t$ versus $t-1$.

\textit{Normalized sentiment} 
\begin{align}
    S^N_{i,t} = \frac{S_{i,t}}{n_{i,t}}
\end{align}
which captures the extent to which all the posts on the forum about asset $i$ at time $t$, $n_{i,t}$, agree on the sentiment. 

\textit{Normalized sentiment change}
\begin{align}
\Delta S^N_{i,t} = S^N_{i,t} - S^N_{i,t-1}
\end{align}
which captures the extent of agreement on the forum around a specific sentiment. 

\paragraph{Results} Table \ref{tab:long-term_signals} presents the results for the relationship between various long-term signals and market variables. We select stocks that have a non-zero number of posts both within the time period $t$ and $t+1$ for our sample. We observe that, consistently with our previous analysis - returns are particularly hard to predict using signals from WSB, but also that stocks within our sample have a tendency to revert in returns, as per the negative coefficient on $r_{i,t}$. The change in momentum has a consistent and negative effect, potentially capturing the reversion in hype, which is known to follow patterns similar to the hump-shaped patterns in infectious disease.  
\subsection{Due dilligence tables}
\label{app:due_dilligence_tables}

We present the comprehensive results of our exploration of due diligence posts and their link to returns. 

\begin{table*}[h]
\begin{adjustbox}{width=\columnwidth,totalheight=\textheight-2\baselineskip,center}
    \begin{tabular}{lllllll}
\toprule
 &  &  & Sentiment & Other Variable & CAR & $r_{[t,t-1]}$ \\
Time Filter & Meme Stocks Excluded & Other Variable Name &  &  &  &  \\
\midrule
All posts & False &  & -0.0102 &  & -0.0356 & -0.0048 \\
 &  &  & (0.0072) &  & (0.011)** & (0.0123) \\
 &  & Num comments & -0.0345 & 0.0113 & -0.0356 & -0.0048 \\
 &  &  & (0.0134)** & (0.0079) & (0.011)** & (0.0123) \\
 &  & Post word count & -0.0301 & 0.0052 & -0.0356 & -0.0048 \\
 &  &  & (0.0099)** & (0.0033) & (0.011)** & (0.0123) \\
 &  & Max comment depth & -0.0316 & 0.0159 & -0.0356 & -0.0048 \\
 &  &  & (0.0232) & (0.0183) & (0.011)** & (0.0123) \\
 &  & Proactive & -0.0404 & 0.0443 & -0.0356 & -0.0048 \\
 &  &  & (0.0121)*** & (0.0119)*** & (0.011)** & (0.0123) \\
 &  & Includes URL & -0.0059 & -0.0116 & -0.0356 & -0.0048 \\
 &  &  & (0.0089) & (0.0112) & (0.011)** & (0.0123) \\
\cline{2-7}
 & True &  & -0.0112 &  & -0.0368 & -0.0049 \\
 &  &  & (0.006) &  & (0.0111)*** & (0.0123) \\
 &  & Num comments & -0.0106 & -0.0002 & -0.0368 & -0.0049 \\
 &  &  & (0.0109) & (0.0038) & (0.0111)*** & (0.0123) \\
 &  & Post word count & 0.0153 & -0.0057 & -0.0368 & -0.0049 \\
 &  &  & (0.0158) & (0.003) & (0.0111)*** & (0.0123) \\
 &  & Max comment depth & -0.0063 & -0.0036 & -0.0368 & -0.0049 \\
 &  &  & (0.0101) & (0.0057) & (0.0111)*** & (0.0123) \\
 &  & Proactive & -0.0389 & 0.0326 & -0.0368 & -0.0049 \\
 &  &  & (0.018)* & (0.0172) & (0.0111)*** & (0.0123) \\
 &  & Includes URL & -0.0094 & -0.0049 & -0.0368 & -0.0049 \\
 &  &  & (0.0066) & (0.0094) & (0.0111)*** & (0.0123) \\
\cline{1-7} \cline{2-7}
Post-2021 & False &  & -0.0304 &  & -0.0358 & -0.005 \\
 &  &  & (0.0139)* &  & (0.011)** & (0.0123) \\
 &  & Num comments & -0.0518 & 0.0144 & -0.0358 & -0.005 \\
 &  &  & (0.0107)*** & (0.0116) & (0.011)** & (0.0123) \\
 &  & Post word count & -0.0414 & 0.0041 & -0.0358 & -0.005 \\
 &  &  & (0.0055)*** & (0.0045) & (0.011)** & (0.0123) \\
 &  & Max comment depth & -0.0926 & 0.0495 & -0.0358 & -0.005 \\
 &  &  & (0.0319)** & (0.0297) & (0.011)** & (0.0123) \\
 &  & Proactive & -0.0502 & 0.0463 & -0.0358 & -0.005 \\
 &  &  & (0.015)*** & (0.0192)* & (0.011)** & (0.0123) \\
 &  & Includes URL & -0.0205 & -0.025 & -0.0358 & -0.005 \\
 &  &  & (0.0224) & (0.0294) & (0.011)** & (0.0123) \\
\cline{2-7}
 & True &  & -0.0439 &  & -0.0373 & -0.005 \\
 &  &  & (0.0112)*** &  & (0.0111)*** & (0.0123) \\
 &  & Num comments & -0.0456 & 0.0007 & -0.0373 & -0.005 \\
 &  &  & (0.0184)* & (0.0068) & (0.0111)*** & (0.0123) \\
 &  & Post word count & -0.0257 & -0.0039 & -0.0373 & -0.005 \\
 &  &  & (0.0351) & (0.0061) & (0.0111)*** & (0.0123) \\
 &  & Max comment depth & -0.0515 & 0.0061 & -0.0373 & -0.005 \\
 &  &  & (0.0173)** & (0.0107) & (0.0111)*** & (0.0123) \\
 &  & Proactive & -0.0807 & 0.0428 & -0.0373 & -0.005 \\
 &  &  & (0.0316)* & (0.0297) & (0.0111)*** & (0.0123) \\
 &  & Includes URL & -0.0477 & 0.0085 & -0.0373 & -0.005 \\
 &  &  & (0.0135)*** & (0.017) & (0.0111)*** & (0.0123) \\
\cline{1-7} \cline{2-7}
Pre-2021 & False &  & 0.0032 &  & -0.0373 & -0.005 \\
 &  &  & (0.009) &  & (0.0112)*** & (0.0123) \\
 &  & Num comments & -0.0019 & 0.002 & -0.0373 & -0.005 \\
 &  &  & (0.013) & (0.0063) & (0.0112)*** & (0.0123) \\
 &  & Post word count & 0.0304 & -0.0059 & -0.0373 & -0.005 \\
 &  &  & (0.0171) & (0.0035) & (0.0112)*** & (0.0123) \\
 &  & Max comment depth & 0.0109 & -0.0055 & -0.0373 & -0.005 \\
 &  &  & (0.0118) & (0.0082) & (0.0112)*** & (0.0123) \\
 &  & Proactive & -0.0158 & 0.0224 & -0.0373 & -0.005 \\
 &  &  & (0.023) & (0.0202) & (0.0112)*** & (0.0123) \\
 &  & Includes URL & 0.0046 & -0.0041 & -0.0373 & -0.005 \\
 &  &  & (0.0092) & (0.0107) & (0.0112)*** & (0.0123) \\
\cline{2-7}
 & True &  & -0.003 &  & -0.037 & -0.0052 \\
 &  &  & (0.0068) &  & (0.0111)*** & (0.0123) \\
 &  & Num comments & 0.0045 & -0.0029 & -0.037 & -0.0052 \\
 &  &  & (0.0121) & (0.0043) & (0.0111)*** & (0.0123) \\
 &  & Post word count & 0.0251 & -0.0061 & -0.037 & -0.0052 \\
 &  &  & (0.0181) & (0.0036) & (0.0111)*** & (0.0123) \\
 &  & Max comment depth & 0.0091 & -0.0087 & -0.037 & -0.0052 \\
 &  &  & (0.0117) & (0.0067) & (0.0111)*** & (0.0123) \\
 &  & Proactive & -0.0292 & 0.0309 & -0.037 & -0.0052 \\
 &  &  & (0.0205) & (0.0197) & (0.0111)*** & (0.0123) \\
 &  & Includes URL & -0.0012 & -0.005 & -0.037 & -0.0052 \\
 &  &  & (0.0073) & (0.0111) & (0.0111)*** & (0.0123) \\
\cline{1-7} \cline{2-7}
\bottomrule
\end{tabular}

\end{adjustbox}
\caption{Panel regression results for the bullish post model where the dependent variable is the asset price change four weeks after the post. Standard errors are in parentheses, with * indicating levels of statistical significance.}
\label{tab:fe_results_bull_4}
\end{table*}

\begin{table*}[h]
\begin{adjustbox}{width=\columnwidth,totalheight=\textheight-2\baselineskip,center}
    \begin{tabular}{lllllll}
\toprule
 &  &  & Sentiment & Other Variable & CAR & $r_{[t,t-1]}$ \\
Time Filter & Meme Stocks Excluded & Other Variable Name &  &  &  &  \\
\midrule
All posts & False &  & -0.0037 &  & -0.0363 & -0.0029 \\
 &  &  & (0.0139) &  & (0.011)*** & (0.012) \\
 &  & Num comments & -0.0039 & -0.0001 & -0.0363 & -0.0029 \\
 &  &  & (0.0204) & (0.0094) & (0.011)*** & (0.012) \\
 &  & Post word count & 0.0234 & -0.0063 & -0.0363 & -0.0029 \\
 &  &  & (0.0247) & (0.0063) & (0.011)*** & (0.012) \\
 &  & Max comment depth & -0.0008 & -0.002 & -0.0363 & -0.0029 \\
 &  &  & (0.0245) & (0.0141) & (0.011)*** & (0.012) \\
 &  & Proactive & 0.0211 & -0.0331 & -0.0363 & -0.0029 \\
 &  &  & (0.0334) & (0.0363) & (0.011)*** & (0.012) \\
 &  & Includes URL & -0.012 & 0.0228 & -0.0363 & -0.0029 \\
 &  &  & (0.0162) & (0.0164) & (0.011)*** & (0.012) \\
\cline{2-7}
 & True &  & 0.0051 &  & -0.037 & -0.003 \\
 &  &  & (0.0125) &  & (0.0111)*** & (0.0121) \\
 &  & Num comments & -0.0076 & -0.0046 & -0.037 & -0.003 \\
 &  &  & (0.0198) & (0.0081) & (0.0111)*** & (0.0121) \\
 &  & Post word count & 0.0248 & -0.0045 & -0.037 & -0.003 \\
 &  &  & (0.0228) & (0.0047) & (0.0111)*** & (0.0121) \\
 &  & Max comment depth & 0.0049 & 0.0001 & -0.037 & -0.003 \\
 &  &  & (0.022) & (0.0136) & (0.0111)*** & (0.0121) \\
 &  & Proactive & 0.0245 & -0.025 & -0.037 & -0.003 \\
 &  &  & (0.0383) & (0.0378) & (0.0111)*** & (0.0121) \\
 &  & Includes URL & 0.0004 & 0.0124 & -0.037 & -0.003 \\
 &  &  & (0.0145) & (0.0168) & (0.0111)*** & (0.0121) \\
\cline{1-7} \cline{2-7}
Post-2021 & False &  & -0.0451 &  & -0.0367 & -0.0055 \\
 &  &  & (0.0354) &  & (0.011)*** & (0.0123) \\
 &  & Num comments & -0.0588 & -0.0055 & -0.0367 & -0.0055 \\
 &  &  & (0.1576) & (0.0596) & (0.011)*** & (0.0123) \\
 &  & Post word count & 0.029 & -0.0197 & -0.0367 & -0.0055 \\
 &  &  & (0.1315) & (0.0342) & (0.011)*** & (0.0123) \\
 &  & Max comment depth & -0.2937 & 0.1583 & -0.0367 & -0.0055 \\
 &  &  & (0.2155) & (0.1456) & (0.011)*** & (0.0123) \\
 &  & Proactive & -0.0293 & -0.0385 & -0.0367 & -0.0055 \\
 &  &  & (0.0721) & (0.1319) & (0.011)*** & (0.0123) \\
 &  & Includes URL & -0.0395 & -0.0171 & -0.0367 & -0.0055 \\
 &  &  & (0.0494) & (0.105) & (0.011)*** & (0.0123) \\
\cline{2-7}
 & True &  & -0.0389 &  & -0.0375 & -0.0056 \\
 &  &  & (0.0569) &  & (0.0111)*** & (0.0123) \\
 &  & Num comments & -0.391 & -0.1153 & -0.0375 & -0.0056 \\
 &  &  & (0.1963)* & (0.0553)* & (0.0111)*** & (0.0123) \\
 &  & Post word count & -0.5337 & 0.0934 & -0.0375 & -0.0056 \\
 &  &  & (0.4621) & (0.0824) & (0.0111)*** & (0.0123) \\
 &  & Max comment depth & -0.1592 & 0.0861 & -0.0375 & -0.0056 \\
 &  &  & (0.1234) & (0.0839) & (0.0111)*** & (0.0123) \\
 &  & Proactive & -0.2465 & 0.2475 & -0.0375 & -0.0056 \\
 &  &  & (0.2097) & (0.2032) & (0.0111)*** & (0.0123) \\
 &  & Includes URL & -0.0532 & 0.0317 & -0.0375 & -0.0056 \\
 &  &  & (0.0712) & (0.111) & (0.0111)*** & (0.0123) \\
\cline{1-7} \cline{2-7}
Pre-2021 & False &  & 0.0001 &  & -0.0371 & -0.0027 \\
 &  &  & (0.0146) &  & (0.0111)*** & (0.0121) \\
 &  & Num comments & 0.0031 & 0.0011 & -0.0371 & -0.0027 \\
 &  &  & (0.0194) & (0.0081) & (0.0111)*** & (0.0121) \\
 &  & Post word count & 0.0226 & -0.0052 & -0.0371 & -0.0027 \\
 &  &  & (0.0229) & (0.0046) & (0.0111)*** & (0.0121) \\
 &  & Max comment depth & 0.0124 & -0.0088 & -0.0371 & -0.0027 \\
 &  &  & (0.0223) & (0.0146) & (0.0111)*** & (0.0121) \\
 &  & Proactive & 0.0334 & -0.0427 & -0.0371 & -0.0027 \\
 &  &  & (0.0387) & (0.0392) & (0.0111)*** & (0.0121) \\
 &  & Includes URL & -0.0094 & 0.0258 & -0.0371 & -0.0027 \\
 &  &  & (0.0188) & (0.0212) & (0.0111)*** & (0.0121) \\
\cline{2-7}
 & True &  & 0.0069 &  & -0.037 & -0.0027 \\
 &  &  & (0.013) &  & (0.0111)*** & (0.0121) \\
 &  & Num comments & 0.0058 & -0.0004 & -0.037 & -0.0027 \\
 &  &  & (0.0192) & (0.0082) & (0.0111)*** & (0.0121) \\
 &  & Post word count & 0.0298 & -0.0053 & -0.037 & -0.0027 \\
 &  &  & (0.0221) & (0.0047) & (0.0111)*** & (0.0121) \\
 &  & Max comment depth & 0.01 & -0.0023 & -0.037 & -0.0027 \\
 &  &  & (0.0225) & (0.0139) & (0.0111)*** & (0.0121) \\
 &  & Proactive & 0.0325 & -0.033 & -0.037 & -0.0027 \\
 &  &  & (0.0392) & (0.0388) & (0.0111)*** & (0.0121) \\
 &  & Includes URL & 0.0024 & 0.0122 & -0.037 & -0.0027 \\
 &  &  & (0.0151) & (0.0172) & (0.0111)*** & (0.0121) \\
\cline{1-7} \cline{2-7}
\bottomrule
\end{tabular}

\end{adjustbox}
    \caption{Panel regression results for the bearish post model where the dependent variable is the asset price change four weeks after the post. Standard errors are in parentheses, with * indicating levels of statistical significance.}
    \label{tab:fe_results_bear_4}
\end{table*}

\begin{table*}[h]
\begin{adjustbox}{width=\columnwidth,totalheight=\textheight-2\baselineskip,center}
    \begin{tabular}{lllllll}
\toprule
 &  &  & Sentiment & Other Variable & CAR & $r_{[t,t-1]}$ \\
Time Filter & Meme Stocks Excluded & Other Variable Name &  &  &  &  \\
\midrule
All posts & False &  & 0.0158 &  & -0.0512 & -0.011 \\
 &  &  & (0.0312) &  & (0.0148)*** & (0.0162) \\
 &  & Num comments & 0.0067 & 0.0042 & -0.0512 & -0.011 \\
 &  &  & (0.0265) & (0.0076) & (0.0148)*** & (0.0162) \\
 &  & Post word count & 0.0672 & -0.0135 & -0.0512 & -0.011 \\
 &  &  & (0.0191)*** & (0.0038)*** & (0.0148)*** & (0.0162) \\
 &  & Max comment depth & -0.0377 & 0.0398 & -0.0512 & -0.011 \\
 &  &  & (0.0181)* & (0.0275) & (0.0148)*** & (0.0162) \\
 &  & Proactive & 0.0033 & 0.0183 & -0.0512 & -0.011 \\
 &  &  & (0.0445) & (0.0218) & (0.0148)*** & (0.0162) \\
 &  & Includes URL & 0.0174 & -0.0043 & -0.0512 & -0.011 \\
 &  &  & (0.0339) & (0.0155) & (0.0148)*** & (0.0162) \\
\cline{2-7}
 & True &  & -0.0224 &  & -0.0513 & -0.0109 \\
 &  &  & (0.0101)* &  & (0.0148)*** & (0.0162) \\
 &  & Num comments & -0.0319 & 0.0038 & -0.0513 & -0.0109 \\
 &  &  & (0.0189) & (0.0074) & (0.0148)*** & (0.0162) \\
 &  & Post word count & 0.0497 & -0.0156 & -0.0513 & -0.0109 \\
 &  &  & (0.0278) & (0.0058)** & (0.0148)*** & (0.0162) \\
 &  & Max comment depth & -0.0383 & 0.0117 & -0.0513 & -0.0109 \\
 &  &  & (0.0146)** & (0.0085) & (0.0148)*** & (0.0162) \\
 &  & Proactive & -0.09 & 0.0793 & -0.0513 & -0.0109 \\
 &  &  & (0.0288)** & (0.0268)** & (0.0148)*** & (0.0162) \\
 &  & Includes URL & -0.025 & 0.0071 & -0.0513 & -0.0109 \\
 &  &  & (0.0115)* & (0.0139) & (0.0148)*** & (0.0162) \\
\cline{1-7} \cline{2-7}
Post-2021 & False &  & 0.0241 &  & -0.0512 & -0.0109 \\
 &  &  & (0.0481) &  & (0.0148)*** & (0.0162) \\
 &  & Num comments & 0.0105 & 0.0092 & -0.0511 & -0.0109 \\
 &  &  & (0.0297) & (0.0223) & (0.0148)*** & (0.0162) \\
 &  & Post word count & 0.0599 & -0.0133 & -0.0512 & -0.0109 \\
 &  &  & (0.0129)*** & (0.0113) & (0.0148)*** & (0.0162) \\
 &  & Max comment depth & -0.0925 & 0.0927 & -0.0511 & -0.0109 \\
 &  &  & (0.0383)* & (0.0511) & (0.0148)*** & (0.0162) \\
 &  & Proactive & 0.0131 & 0.0257 & -0.0512 & -0.0109 \\
 &  &  & (0.0478) & (0.0286) & (0.0148)*** & (0.0162) \\
 &  & Includes URL & 0.0354 & -0.0284 & -0.0511 & -0.0109 \\
 &  &  & (0.0624) & (0.0492) & (0.0148)*** & (0.0162) \\
\cline{2-7}
 & True &  & -0.0795 &  & -0.0515 & -0.0104 \\
 &  &  & (0.017)*** &  & (0.0148)*** & (0.0161) \\
 &  & Num comments & -0.0852 & 0.0026 & -0.0515 & -0.0104 \\
 &  &  & (0.0314)** & (0.0127) & (0.0148)*** & (0.0161) \\
 &  & Post word count & -0.0112 & -0.0147 & -0.0515 & -0.0104 \\
 &  &  & (0.0539) & (0.0111) & (0.0148)*** & (0.0161) \\
 &  & Max comment depth & -0.1014 & 0.0175 & -0.0515 & -0.0104 \\
 &  &  & (0.0289)*** & (0.0189) & (0.0148)*** & (0.0161) \\
 &  & Proactive & -0.2773 & 0.2298 & -0.0515 & -0.0104 \\
 &  &  & (0.0591)*** & (0.0607)*** & (0.0148)*** & (0.0161) \\
 &  & Includes URL & -0.1023 & 0.0509 & -0.0515 & -0.0104 \\
 &  &  & (0.0218)*** & (0.0291) & (0.0148)*** & (0.0161) \\
\cline{1-7} \cline{2-7}
Pre-2021 & False &  & 0.0102 &  & -0.0522 & -0.0107 \\
 &  &  & (0.02) &  & (0.0148)*** & (0.0162) \\
 &  & Num comments & -0.0057 & 0.0061 & -0.0522 & -0.0107 \\
 &  &  & (0.0206) & (0.0096) & (0.0148)*** & (0.0162) \\
 &  & Post word count & 0.1025 & -0.0202 & -0.0522 & -0.0107 \\
 &  &  & (0.0447)* & (0.0072)** & (0.0148)*** & (0.0162) \\
 &  & Max comment depth & -0.011 & 0.0151 & -0.0522 & -0.0107 \\
 &  &  & (0.0158) & (0.0132) & (0.0148)*** & (0.0162) \\
 &  & Proactive & -0.0212 & 0.037 & -0.0522 & -0.0107 \\
 &  &  & (0.0332) & (0.0248) & (0.0148)*** & (0.0162) \\
 &  & Includes URL & 0.009 & 0.0034 & -0.0522 & -0.0107 \\
 &  &  & (0.0194) & (0.0153) & (0.0148)*** & (0.0162) \\
\cline{2-7}
 & True &  & -0.0081 &  & -0.0519 & -0.0111 \\
 &  &  & (0.0116) &  & (0.0148)*** & (0.0162) \\
 &  & Num comments & -0.009 & 0.0004 & -0.0519 & -0.0111 \\
 &  &  & (0.0209) & (0.0083) & (0.0148)*** & (0.0162) \\
 &  & Post word count & 0.0629 & -0.0154 & -0.0519 & -0.0111 \\
 &  &  & (0.0299)* & (0.0061)* & (0.0148)*** & (0.0162) \\
 &  & Max comment depth & -0.0163 & 0.0058 & -0.0519 & -0.0111 \\
 &  &  & (0.0156) & (0.0096) & (0.0148)*** & (0.0162) \\
 &  & Proactive & -0.0465 & 0.0452 & -0.0519 & -0.0111 \\
 &  &  & (0.0281) & (0.0264) & (0.0148)*** & (0.0162) \\
 &  & Includes URL & -0.0086 & 0.0014 & -0.0519 & -0.0111 \\
 &  &  & (0.0125) & (0.0155) & (0.0148)*** & (0.0162) \\
\cline{1-7} \cline{2-7}
\bottomrule
\end{tabular}

\end{adjustbox}
    \caption{Panel regression results for the bullish post model where the dependent variable is the asset price change twelve weeks after the post. Standard errors are in parentheses, with * indicating levels of statistical significance.}
    \label{tab:fe_results_bull_12}
\end{table*}

\begin{table*}[h]
\begin{adjustbox}{width=\columnwidth,totalheight=\textheight-2\baselineskip,center}
    \begin{tabular}{lllllll}
\toprule
 &  &  & Sentiment & Other Variable & CAR & $r_{[t,t-1]}$ \\
Time Filter & Meme Stocks Excluded & Other Variable Name &  &  &  &  \\
\midrule
All posts & False &  & -0.0098 &  & -0.0513 & -0.0084 \\
 &  &  & (0.0201) &  & (0.0148)*** & (0.0159) \\
 &  & Num comments & -0.018 & -0.003 & -0.0513 & -0.0084 \\
 &  &  & (0.0226) & (0.0111) & (0.0148)*** & (0.0159) \\
 &  & Post word count & 0.0488 & -0.0137 & -0.0513 & -0.0084 \\
 &  &  & (0.0298) & (0.0076) & (0.0148)*** & (0.0159) \\
 &  & Max comment depth & -0.0219 & 0.0086 & -0.0513 & -0.0084 \\
 &  &  & (0.0279) & (0.0139) & (0.0148)*** & (0.0159) \\
 &  & Proactive & 0.0165 & -0.0351 & -0.0513 & -0.0084 \\
 &  &  & (0.0344) & (0.0318) & (0.0148)*** & (0.0159) \\
 &  & Includes URL & -0.0309 & 0.058 & -0.0513 & -0.0084 \\
 &  &  & (0.0234) & (0.0234)* & (0.0148)*** & (0.0159) \\
\cline{2-7}
 & True &  & 0.0012 &  & -0.0516 & -0.0087 \\
 &  &  & (0.0195) &  & (0.0148)*** & (0.0159) \\
 &  & Num comments & -0.0343 & -0.0129 & -0.0516 & -0.0087 \\
 &  &  & (0.0217) & (0.0093) & (0.0148)*** & (0.0159) \\
 &  & Post word count & 0.0414 & -0.0093 & -0.0516 & -0.0087 \\
 &  &  & (0.0338) & (0.0073) & (0.0148)*** & (0.0159) \\
 &  & Max comment depth & -0.0172 & 0.0132 & -0.0516 & -0.0087 \\
 &  &  & (0.0276) & (0.0141) & (0.0148)*** & (0.0159) \\
 &  & Proactive & 0.0354 & -0.0439 & -0.0516 & -0.0087 \\
 &  &  & (0.0383) & (0.0363) & (0.0148)*** & (0.0159) \\
 &  & Includes URL & -0.0151 & 0.0438 & -0.0516 & -0.0087 \\
 &  &  & (0.0222) & (0.0227) & (0.0148)*** & (0.0159) \\
\cline{1-7} \cline{2-7}
Post-2021 & False &  & -0.093 &  & -0.0518 & -0.0106 \\
 &  &  & (0.0445)* &  & (0.0148)*** & (0.0161) \\
 &  & Num comments & 0.1031 & 0.0793 & -0.0518 & -0.0105 \\
 &  &  & (0.1535) & (0.0858) & (0.0148)*** & (0.0161) \\
 &  & Post word count & 0.141 & -0.0622 & -0.0518 & -0.0106 \\
 &  &  & (0.0991) & (0.0416) & (0.0148)*** & (0.0161) \\
 &  & Max comment depth & -0.1093 & 0.0104 & -0.0518 & -0.0106 \\
 &  &  & (0.2298) & (0.1442) & (0.0148)*** & (0.0161) \\
 &  & Proactive & -0.0423 & -0.1243 & -0.0518 & -0.0106 \\
 &  &  & (0.0322) & (0.0988) & (0.0148)*** & (0.0161) \\
\cline{2-7}
 & True &  & -0.0005 &  & -0.0521 & -0.0108 \\
 &  &  & (0.0729) &  & (0.0148)*** & (0.0162) \\
 &  & Num comments & -0.2574 & -0.0841 & -0.0521 & -0.0108 \\
 &  &  & (0.1374) & (0.0542) & (0.0148)*** & (0.0162) \\
 &  & Post word count & 0.1286 & -0.0244 & -0.0521 & -0.0108 \\
 &  &  & (0.2413) & (0.0478) & (0.0148)*** & (0.0162) \\
 &  & Max comment depth & 0.0101 & -0.0076 & -0.0521 & -0.0108 \\
 &  &  & (0.1246) & (0.1226) & (0.0148)*** & (0.0162) \\
 &  & Proactive & 0.0893 & -0.107 & -0.0521 & -0.0108 \\
 &  &  & (0.0894) & (0.063) & (0.0148)*** & (0.0162) \\
 &  & Includes URL & -0.0493 & 0.1081 & -0.0521 & -0.0108 \\
 &  &  & (0.0492) & (0.1519) & (0.0148)*** & (0.0162) \\
\cline{1-7} \cline{2-7}
Pre-2021 & False &  & -0.0021 &  & -0.0517 & -0.0084 \\
 &  &  & (0.0201) &  & (0.0148)*** & (0.0159) \\
 &  & Num comments & -0.0282 & -0.0095 & -0.0517 & -0.0084 \\
 &  &  & (0.0215) & (0.0092) & (0.0148)*** & (0.0159) \\
 &  & Post word count & 0.0313 & -0.0078 & -0.0517 & -0.0084 \\
 &  &  & (0.0348) & (0.0074) & (0.0148)*** & (0.0159) \\
 &  & Max comment depth & -0.0187 & 0.0119 & -0.0517 & -0.0085 \\
 &  &  & (0.0274) & (0.014) & (0.0148)*** & (0.0159) \\
 &  & Proactive & 0.0309 & -0.0424 & -0.0517 & -0.0084 \\
 &  &  & (0.0391) & (0.0377) & (0.0148)*** & (0.0159) \\
 &  & Includes URL & -0.0242 & 0.0597 & -0.0517 & -0.0085 \\
 &  &  & (0.0239) & (0.0254)* & (0.0148)*** & (0.0159) \\
\cline{2-7}
 & True &  & 0.0013 &  & -0.0516 & -0.0085 \\
 &  &  & (0.0202) &  & (0.0148)*** & (0.0159) \\
 &  & Num comments & -0.026 & -0.01 & -0.0516 & -0.0085 \\
 &  &  & (0.0216) & (0.0094) & (0.0148)*** & (0.0159) \\
 &  & Post word count & 0.0407 & -0.0092 & -0.0516 & -0.0085 \\
 &  &  & (0.0335) & (0.0074) & (0.0148)*** & (0.0159) \\
 &  & Max comment depth & -0.018 & 0.0138 & -0.0516 & -0.0085 \\
 &  &  & (0.0284) & (0.0139) & (0.0148)*** & (0.0159) \\
 &  & Proactive & 0.0338 & -0.0419 & -0.0516 & -0.0085 \\
 &  &  & (0.0394) & (0.0375) & (0.0148)*** & (0.0159) \\
 &  & Includes URL & -0.0138 & 0.0409 & -0.0516 & -0.0086 \\
 &  &  & (0.0231) & (0.0225) & (0.0148)*** & (0.0159) \\
\cline{1-7} \cline{2-7}
\bottomrule
\end{tabular}

\end{adjustbox}
    \caption{Panel regression results for the bearish post model where the dependent variable is the asset price change twelve weeks after the post. Standard errors are in parentheses, with * indicating levels of statistical significance.}
    \label{tab:fe_results_bear_12}
\end{table*}


\end{document}